\documentclass[prd,showpacs,aps,prd,nofootinbib,floatfix,amsmath,amssymb]{revtex4}
\usepackage{graphicx}
\usepackage{slashed}
\usepackage{textcomp}
\begin{document}

\makeatletter
%Feynman slash
\newbox\slashbox \setbox\slashbox=\hbox{$/$}
\newbox\Slashbox \setbox\Slashbox=\hbox{\large$/$}
\def\pFMslash#1{\setbox\@tempboxa=\hbox{$#1$}
  \@tempdima=0.5\wd\slashbox \advance\@tempdima 0.5\wd\@tempboxa
  \copy\slashbox \kern-\@tempdima \box\@tempboxa}
\def\pFMSlash#1{\setbox\@tempboxa=\hbox{$#1$}
  \@tempdima=0.5\wd\Slashbox \advance\@tempdima 0.5\wd\@tempboxa
  \copy\Slashbox \kern-\@tempdima \box\@tempboxa}
\def\FMslash{\protect\pFMslash}
\def\FMSlash{\protect\pFMSlash}
\def\miss#1{\ifmmode{/\mkern-11mu #1}\else{${/\mkern-11mu #1}$}\fi}
%%%% Uso:  \pFMSlash{p}
\makeatother

%\tightenli
\title{Interplay between neutrino magnetic moments and $CP$ violating phases in left--right models}

\author{ D. Delepine$^{a}$, and H. Novales--S\'anchez$^{b}$}

\address{$^a$Divisi\'on de Ciencias e Ingenier\'ias, Universidad de Guanajuato Campus Le\'on, Loma del Bosque 103, Colonia Lomas del Campestre, 37150, Le\'on, Guanajuato, M\'exico.
\\
$^b$Facultad de Ciencias F\'isico Matem\'aticas, Benem\'erita Universidad Aut\'onoma de Puebla, Apartado Postal 1152, Puebla, Puebla, M\'exico.}

\begin{abstract}
We revisit the neutrino magnetic moments (MMs)  in the left--right model with non--manifest symmetry. After deriving an expression in terms of the Dirac and Majorana phases, we analyze the sensitivity of neutrino MMs to these $CP$--violating phases in two scenarios: 1) a maximal right mixing in which left-- and right--handed neutrinos are mixed by the same matrix; and 2) a right--handed neutrino mixing whose off--diagonal entries are much smaller than the elements in the diagonal, but where  the $CP$ phases remain general. Our results show that, even though certain values of the Majorana phases can eliminate neutrino MMs, the presence of a maximal $CP$--violating phase in neutrino mixing matrix, as favored by the discrepancy between T2K results and reactor measurements in neutrino oscillations, requires that at least one neutrino have a large nonzero MM.
\end{abstract}


\maketitle

\section{Introduction}
\label{intro}
The detection~\cite{HatCERN} of a Higgs--like particle with mass around $125\,{\rm GeV}$ at the CERN Large Hadron Collider has been a quite significant discovery. Though, so far, analyses of experimental data by the ATLAS~\cite{SMHatATLAS} and CMS~\cite{SMHatCMS} Collaborations on such scalar particle are consistent with the Standard Model (SM) candidate, which points towards the conclusion of this beautiful low--energy description of nature, the community is convinced that extension of such model is imperious. It is currently accepted that the SM has theoretical problems which call for a solution. Even more important, there is clear experimental evidence of physical phenomena that remain outside the reach of the SM: no role is played by the gravitational interaction; no explanation is provided for the existence of dark matter nor dark energy; the neutrinos are massive.
\\

When the existence of neutrinos was first proposed by Pauli~\cite{WP}, their properties, at the time considered exotic and very challenging to measure, included the possibility that these particles were massive. The remarkable idea that neutrino flavors oscillate with distance~\cite{BP,GK} has driven trends of research in neutrino physics. The most important reason for this is, perhaps, that the main interpretation of neutrino oscillations requires neutrinos to be massive in order for such a phenomenon to take place\footnote{Other sorts of new physics, such as violation of Lorentz invariance, provide alternative interpretations~\cite{LVnuosc} of neutrino oscillations.}. The already accomplished measurement~\cite{nuoscmsmnt} of the last neutrino mixing angle has supplied strong evidence supporting this idea.
\\

Since neutrinos are neutral and massive fermions, they can be represented by either Dirac or Majorana fields~\cite{GK,EMaj}. This interesting possibility incarnates one of the most relevant questions about neutrino physics at present. An answer has been looked for by several experimental groups that aim at observing the rare neutrinoless double beta decay ($2\beta_{0\nu}$), which can occur only in the presence of Majorana neutrinos. Experimental setups are based on the double beta decay of diverse isotopes: $^{76}$Ge~\cite{76Ge} (Hiedelberg--Moscow, IGEX, GERDA), $^{100}$Mo and $^{82}$Se~\cite{100Mo82Se} (NEMO), $^{130}$Te~\cite{130Te} (CUORICINO), $^{150}$Nd~\cite{150Nd} (NEMO), and $^{136}$Xe~\cite{136Xe} (KamLAND--Zen, EXO).
%Other important physical phenomena, as neutrino oscillations, are not able to provide information about the nature of neutrinos.
One important difference among the Dirac and Majorana cases lies on the number of $CP$--violating phases involved in the parametrization of neutrino mixing.
%, for there is only one of them if neutrinos are Dirac--like, whereas Majorana neutrinos require two more phases.
Sensitivity of physical phenomena to these $CP$--violating phases may provide information on the nature of neutrinos, so the analysis of observables depending on them is relevant and interesting.
%The possibility that observables incorporate effects of new physics through these phases is interesting and works on the matter exist~\cite{wonnuCPvph}.
\\

Investigations centered in the electromagnetic properties of neutrinos~\cite{BGS} constitute an active topic in neutrino physics.
Works concerning the neutrino anapole moment~\cite{RMH}, charge radius~\cite{nucr}, and electric dipole moment~\cite{GrNeu,ADD,DvSt1} exist, but most studies have explored the neutrino magnetic moments (MMs). 
This quantity has been calculated in gauge theories~\cite{mmgth} and in a model--independent manner~\cite{BGRMVW} as well. Nice and detailed analyses of these MMs, including the full structure of the neutrino electromagnetic vertex and the issue of gauge independence, were carried out in Refs.~\cite{DvSt1,DvSt2}, within the so--called {\it Minimally Extended Standard Model}. Mechanisms to produce large contributions to neutrino MMs have been propounded~\cite{PtzMo}.
Astrophysical approaches~\cite{nuAsphy} have been quite relevant in this branch of physics.
In the SM, neutrinos are Dirac massless particles that can only have one electromagnetic property~\cite{NRT}, namely, the anapole moment. In general, the introduction of massive neutrinos changes the game rules, opening~\cite{BGS} the possibility of encountering more electromagnetic properties, which are very different~\cite{empDvsM} depending on whether neutrinos are Dirac or Majorana particles. Indeed, it has been claimed~\cite{BGRMVW,BCRVW} that the measurement of a neutrino magnetic moment larger than or equal to $10^{-15}\mu_{\rm B}$ would indicate that neutrinos are Majorana particles. The simplest SM extension that yields neutrino masses is achieved just by adding the right--handed components of the neutrino fields. The prediction of this minimal extension for the neutrino magnetic moment is about eight orders of magnitude smaller than the best upper limits reached by experiments so far, which renders this quantity attractive to look for new physics.
\\

Motivated by this, we direct the attention of the present work to the MMs of neutrinos.
A popular and well known SM extension that yields large contributions to the MMs of Majorana neutrinos is based on the gauge group SU(2)$_L\times$SU(2)$_R\times$U(1)$_{B-L}~$\cite{lrmod}, which is broken down to SU(2)$_L\times$U(1)$_Y$ and then to the electromagnetic gauge group through different stages of spontaneous symmetry breaking. These models, best known as {\it left--right}, give rise to a new heavy charged gauge boson, $W_R$, which comes along with a rich phenomenology.
The production of $W_R$ as a mean to test the Majorana nature of neutrinos was discussed in Ref.~\cite{KS}. It has been shown that the presence of such new boson would have effects on various physical phenomena, such as the neutron electric dipole moment~\cite{EGN}, decays of the Higgs boson~\cite{MPT}, the forward--backward asymmetry~\cite{BKY}, and $K$~\cite{BBPR,MNNS}, $B$~\cite{BFY,BMN} and $D$~\cite{BDGL} physics. The possible discovery of the $W_R$ at the Large Hadron Collider has also been discussed~\cite{Fetal,GKKM}, and its role in charged Higgs production~\cite{ABF} and Lepton Flavor Violation~\cite{DDKV} at this collider has been explored. The decay of this heavy boson to SM gauge bosons has been analyzed as a probe of the symmetry breaking sector~\cite{CLK}. A model--independent global analysis concerning new charged gauge bosons, including this one, can be found in Ref~\cite{HSYY}.
\\

In this paper, we consider a left--right model with non--manifest left--right symmetry and which is endowed with two scalar triplets and one scalar bidoublet defining the Higgs sector. We assume that the right (heavy) charged gauge boson $W_R$ mixes with the left--handed $W_L$, with such mixture being parametrized by an angle $\zeta$.
A calculation of the contributions from this model to neutrino MMs~\cite{CGZ}, carried out in the unitary gauge, showed that the leading term, which does not involve the mass of light neutrinos, yields contributions as large as $10^{-11}\mu_{\rm B}$. In the present paper, we compute the contributions from heavy gauge bosons in the $R_\xi$--gauge and then take the unitary gauge, finding agreement with the results of Ref.~\cite{CGZ}.
Then we express the MMs in terms of the $CP$--violating phases and analyze the sensitiveness of this electromagnetic property to them, with special focus on the Majorana phases. First, we carry out our analysis under the assumption that the mixing of right--handed neutrinos is realized by the same matrix that mixes left--handed neutrinos. This is a particular case of maximal right--handed neutrino mixing. After that, we examine a scenario defined by a less democratic neutrino mixing, which involves a mixing matrix that we assume to be close to identity, except for the $CP$--violating matrices, which we keep general. In both cases, we find that, while for each neutrino there are sets of nonzero specific values of $CP$ phases that cancel its MM, at least one of the three neutrinos must have a nonzero MM when neutrino mixing violates $CP$ invariance. 
\\

We have organized the paper in the following way: in Section~\ref{lrt}, we specify our setup and provide the necessary information to perform the calculation. The one--loop $\nu\nu\gamma$ vertex is calculated in Section~\ref{namm} and the resulting contribution to the MM is presented. In Section~\ref{discussion} we analyze our result, carrying out the discussion around the role played by the Majorana phases. We also show estimations in this section. Finally, we present our conclusions in Section~\ref{conc}.

\section{The model}
\label{lrt}
The essential ingredient of left--right models is the gauge group SU(2)$_L\times$SU(2)$_R\times$U(1)$_{B-L}$, which is a simple extension of the electroweak SM. The enlargement of the SM gauge group by the SU(2)$_R$ introduces a new set of gauge fields, although the fermionic content of the SM remains, in principle, intact. The presence of the new gauge degrees of freedom then allows one to define charged--current interactions for right--handed fermions, which include right--handed neutrinos.
\\

The set of SU(2)$_R$ degrees of freedom is represented by three gauge bosons. The Higgs sector of the left--right model considered in the present paper includes a left triplet, $\Delta_L$, and a right triplet, $\Delta_R$. These multiplets can be represented by the $2\times2$ matrices
\begin{equation}
\Delta_L=
\left(
\begin{array}{cc}
\delta^+_L/\sqrt{2} & \delta^{++}_L
\\ \\
\delta^0_L & -\delta^+_L/\sqrt{2}
\end{array}
\right), \hspace{1cm}
\Delta_R=
\left(
\begin{array}{cc}
\delta^+_R/\sqrt{2} & \delta^{++}_R
\\ \\
\delta^0_R & -\delta^+_R/\sqrt{2}
\end{array}
\right),
\end{equation}
in which doubly charged, singly charged and neutral scalars, respectively represented by $\delta_{L,R}^{++}$, $\delta_{L,R}^+$ and $\delta^0_{L,R}$, are involved. The  interactions of these scalars with the SU(2)$_L$ and SU(2)$_R$ gauge bosons take place through the covariant derivatives
\begin{eqnarray}
D^\mu\Delta_L=\partial^\mu\Delta_L-i\frac{g_L}{2}\left[ W^\mu_L,\Delta_L \right]-igB^\mu\Delta_L,
\\ \nonumber \\
D^\mu\Delta_R=\partial^\mu\Delta_R-i\frac{g_R}{2}\left[ W^\mu_R,\Delta_R \right]-igB^\mu\Delta_R,
\end{eqnarray}
in which $g_L$ is the SU(2)$_L$ coupling, $g_R$ is the SU(2)$_R$ coupling, $g$ is the U(1)$_{B-L}$ coupling, $W^\mu_L=W^{a\mu}_L\,\tau^a/2$, and $W^\mu_R=W^{a\mu}_R\,\tau^a/2$. Here, the $\tau^a$ is a Pauli matrix. In order to couple left-- and right--handed fermionic doublets, and generate fermionic masses, the scalar sector of this model has another multiplet structure, which is a Higgs bidoublet, $\Phi$. The bidoublet $\Phi$ is a $2\times2$ matrix,
\begin{equation}
\Phi=
\left(
\begin{array}{cc}
\phi^0_1 & \phi_1^+
\\
\phi^-_2 & \phi^0_2
\end{array}
\right),
\end{equation}
transforming as $\Phi\to U_L\Phi \,U_R$ under the full gauge group. The bidoublet interacts only with the SU(2)$_L$ and SU(2)$_R$ gauge bosons of the theory. The corresponding interactions are defined by the covariant derivative,
\begin{equation}
D^\mu\Phi=\partial^\mu\Phi-i\left( g_LW^\mu_L\Phi-g_R\Phi W^\mu_R \right).
\end{equation}
\\

In a first stage of spontaneous symmetry breaking, the gauge group SU(2)$_R\times$U(1)$_{B-L}$ is broken down to U(1)$_Y$, so that the remaining gauge group corresponds to the electroweak SM. This is achieved when the $\Delta_R$ triplet acquires the vacuum expectation value
\begin{equation}
\langle \Delta_R \rangle=
\left(
\begin{array}{cc}
0 & 0
\\
v_R & 0
\end{array}
\right).
\end{equation}
This stage of symmetry breaking generates the masses of two charged gauge bosons, $W_R^+$ and $W_R^-$, defined as
\begin{eqnarray}
W^{+\mu}_R&=&\frac{1}{\sqrt{2}}\left( W^{1\mu}_R- iW^{2\mu}_R \right),
\\
W^{-\mu}_R&=&\frac{1}{\sqrt{2}}\left( W^{1\mu}_R+ iW^{2\mu}_R \right),
\end{eqnarray}
and the mass of one neutral boson, $Z'_\mu$, as well, while leaving a massless U(1)$_Y$ gauge boson.
\\

A second symmetry breaking occurs when the Higgs multiplets $\Phi$ and $\Delta_L$ develop the vacuum expectation values
\begin{equation}
\langle \Phi \rangle=
\left(
\begin{array}{cc}
\kappa & 0
\\
0 & \kappa'
\end{array}
\right), \hspace{1cm}
\langle \Delta_L \rangle=
\left(
\begin{array}{cc}
0 & 0
\\
v_L & 0
\end{array}
\right),
\end{equation}
where we allow $\kappa$ and $\kappa'$ to be complex numbers. The symmetry breaking produces bilinear terms in the renormalizable scalar sector, including mixings among the $W^{3\mu}_L$ and the U(1)$_Y$ gauge boson. An outcome of this is the definition of the massive $Z_\mu$ boson and the photon field, $A_\mu$. On the other hand, the mass eigenstates
\begin{eqnarray}
W^{+\mu}_L&=&\frac{1}{\sqrt{2}}\left( W^{1\mu}_L- iW^{2\mu}_L \right),
\\
W^{-\mu}_L&=&\frac{1}{\sqrt{2}}\left( W^{1\mu}_L+ iW^{2\mu}_L \right),
\end{eqnarray}
are defined.
\\

As we mentioned above, the covariant derivative for the bidoublet $\Phi$ defines interactions of this multiplet with both the SU(2)$_L$ and the SU(2)$_R$ gauge bosons. The renormalizable interactions of the bidoublet then induce a bilinear mixing of the left and right charged bosons:
\begin{equation}
\Big( W^{+\mu}_L\,\,\,W^{+\mu}_R \Big)
\left(
\begin{array}{cc}
\frac{1}{2}\left( v_L^2+|\kappa|^2+|\kappa'|^2 \right)g_L^2 & -g_Lg_R|\kappa\,\kappa'|e^{i\omega}
\\ \\
-g_Lg_R|\kappa\,\kappa'|e^{-i\omega} & \frac{1}{2}\left( v_R^2+|\kappa|^2+|\kappa'|^2 \right)g_R^2
\end{array}
\right)
\left(
\begin{array}{c}
W^-_{L\,\mu}
\\ \\
W^-_{R\,\mu}
\end{array}
\right),
\end{equation}
with $\omega$ being the complex phase of the product $\kappa^*\kappa'$. Diagonalization of this mixing matrix leads to the mass eigenstates $W_1$ and $W_2$, given by
\begin{eqnarray}
W^{+\mu}_L&=&\cos\zeta\, W^{+\mu}_1-\sin\zeta\, W^{+\mu}_2,
\label{WLmix}
\\ \nonumber \\
W^{+\mu}_R&=&e^{i\omega}\left( \sin\zeta\,W^{+\mu}_1+\cos\zeta\,W^{+\mu}_2 \right).
\label{WRmix}
\end{eqnarray}
For an SU(2)$_R$ breaking scale such that $v_R\gg|\kappa|,|\kappa'|,v_L$, the mixing angle $\zeta$ and the masses of the $W_1$ and $W_2$ bosons can be expressed as
\begin{eqnarray}
\zeta&\simeq&\frac{g_L}{g_R}\frac{2|\kappa\,\kappa'|}{v_R^2},
\label{zetaapprox}
\\ \nonumber \\
m_{W_1}^2&\simeq&\frac{g_L^2}{2}\left( v_L^2+|\kappa|^2+|\kappa'|^2 \right),
\\ \nonumber \\
m_{W_2}^2&\simeq&\frac{g_R^2}{2}v_R^2.
\end{eqnarray}
The simple expression for the $\zeta$ angle, given in Eq.~(\ref{zetaapprox}), shows that in the limit $v_R\to\infty$, the mixing of the $W_L$ and $W_R$ bosons, Eqs.~(\ref{WLmix}) and (\ref{WRmix}), vanishes, so that the $W_1$ boson coincide with the $W_L$, while the $W_2$ is essentially the $W_R$ charged gauge boson. Note that the $W_2$ mass is proportional to the $v_R$ scale.
\\

One of the first great incentives behind left--right models lied on the capability of some of its versions to incorporate~\cite{MohSen} parity as a fundamental symmetry of nature, while describing a world that does not distinguishes chirality. In these models, the gauge group SU(2)$_L\times$SU(2)$_R\times$U(1)$_{B-L}$ is broken into SU(2)$_L\times$U(1)$_Y$ at the $v_R$ scale, entailing, as a collateral effect, breaking of parity symmetry, which thus provides an elegant explanation of why left and right states couple differently at the electroweak scale. In order for this mechanism to work, a discrete symmetry of the theory under the interchange of left and right states is assumed, which implies the important relation $g_L=g_R$. This assumption, which characterizes models with manifest left--right symmetry, leads to strong constraints on the mass of the heavy charged boson, which is found to be in the TeV range. Indeed, the analysis of the first data provided by the Large Hadron Collider was used recently to establish that $m_{W_R}\gtrsim1.4$\,TeV for masses of right--handed neutrinos of order of a few GeV~\cite{NNSZ}. Mass differences of neutral kaon and $B$--meson, on the other hand, have been employed to derive the even more stringent bound $m_{W_R}>2.5$\,TeV~\cite{ZAJM}. Concerning the mixing angle $\zeta$, the upper limit arising within the manifest left--right symmetry scenario is $\zeta<0.005$~\cite{Wolfs}.
\\

These constraints can be relaxed in more general scenarios. An alternative type of left--right models involves non--manifest left--right symmetry. It is assumed that there is a discrete left--right symmetry, but it is broken at a much higher scale than the distinctive scale of the gauge group SU(2)$_L\times$SU(2)$_R\times$U(1)$_{B-L}$. Within such framework, the couplings $g_L$ and $g_R$ can be different. An appealing feature of models with non--manifest left--right symmetry is that the bounds on the charged boson mass and on the mixing angle $\zeta$ are relaxed. It has been shown that in such context the $W_R$ boson mass can be as light as 0.3\,TeV\cite{OnessEbel} and the mixing as large as $\zeta<0.02$\cite{LgckrSkr}. Besides yielding more relaxed bounds, non--manifest left--right symmetry is interesting because it is an efficacious mean to enhance $CP$--violating effects through the $W_L-W_R$ mixing\cite{DFR1,DFR2}.
\\

Interactions of gauge bosons with leptons take place within the fermionic kinetic sector of the left--right model. Contrastingly to what happens in the SM, the fermionic fields in left--right models are all arranged in left-- and right--handed doublets, which couple, respectively, to the SU(2)$_L$ and SU(2)$_R$ gauge bosons. Such interactions include, by construction, right--handed neutrinos. The leptonic doublets are  given as
\begin{equation}
L^i_L=
\left(
\begin{array}{c}
\nu_{i}
\\
l_i
\end{array}
\right)_L,
\hspace{0.4cm}
L^i_R=
\left(
\begin{array}{c}
\nu_i
\\
l_i
\end{array}
\right)_R,
\end{equation}
where $\nu_i=\nu_e,\nu_\mu,\nu_\tau$ and $l_i=l_e,\,l_\mu,\,l_\tau$ (charged lepton fields). The gauge couplings of the left-- and right--handed leptons, generically denoted by $l$, are introduced through the covariant derivatives
\begin{eqnarray}
D^\mu f_L&=&\partial^\mu f_L-\frac{i}{2}\left( g_L\,\tau^aW^{a\mu}_L+gYB^\mu \right) f_L,
\\ \nonumber \\
D^\mu f_R&=&\partial^\mu f_R-\frac{i}{2}\left( g_R\,\tau^aW^{a\mu}_R+gYB^\mu \right) f_R,
\end{eqnarray}
defined by the charge assignments $L_L:(0,1/2,-1)$ and $L_R:(1/2,0,-1)$.
\\

Neutrino oscillations, first proposed by Bruno Pontecorvo~\cite{BP}, is a quantum phenomenon that consists in the existence of nonzero transition probabilities that neutrinos created with certain flavor be measured, after traveling some distance, as neutrinos with different flavors. In order for this to happen, neutrino oscillations require a mixing among neutrino flavors. This mixing, which connects neutrino flavor states $\nu_e$, $\nu_\mu$, $\nu_\tau$ to mass eigenstates $\nu_1$, $\nu_2$, $\nu_3$, is characterized by the unitary transformation
\begin{equation}
\nu_j=\sum_{\alpha=1,2,3}U_{j\alpha}\nu_\alpha,
\end{equation}
where $U_{j\alpha}$ is the Pontecorvo--Maki--Nakagawa--Sakata (PMNS) mixing matrix. A convenient parametrization~\cite{GK,Petcov} of the PMNS matrix is given in terms of three angles and, depending on whether the massive neutrinos are Dirac or Majorana fermions, of one Dirac or one Dirac and two Majorana $CP$--violating phases. Since the neutrino flavor transition probabilities depend on differences of neutrino masses, the occurrence of this effect implies that neutrinos are massive. The relatively recent measurement~\cite{nuoscmsmnt} of the last mixing angle has elevated the status of this interesting conclusion to a fact of great relevance.
\\

The description of flavor neutrino mixing by means of the PMNS matrix introduces $CP$ violation, similarly to what happens in the case of the Cabibbo--Kobayashi--Maskawa (CKM) mixing matrix, nested in the quark sector of the SM. Violations of $CP$ invariance in the lepton sector are driven by the complex phases that are present in such parametrization of the PMNS matrix. Neutrino oscillations can be used to investigate $CP$ violation introduced by the complex character of the PMNS matrix. Transition probabilities $P(\nu_k\to\nu_j)$ and $P(\bar{\nu}_k\to\bar{\nu}_j)$, among different neutrino flavors, $\nu_k$ and $\nu_j$, and antineutrino flavors, $\bar{\nu}_k$ and $\bar{\nu}_j$, involve quartic products of PMNS matrix elements that look like $U^*_{j\alpha}U_{k\alpha}U_{j\beta}U^*_{j\beta}$, which are invariant under the rephasing transformations $U_{j\alpha}\to e^{i\psi_j}U_{j\alpha}\,e^{i\phi_\alpha}$. The Majorana $CP$--violating phases, which exist in the PMNS matrix only if neutrinos are Majorana particles, can be factored out, leaving a CKM--like matrix, involving only the Dirac phase, and a diagonal matrix whose nonzero entries are Majorana phases and 1. Invariance under the rephasings that we just commented yields the cancellation of all Majorana phases in the transition probabilities between neutrino flavors and between antineutrino flavors as well, so that $CP$--violating effects due to Majorana phases are~\cite{GK,BHP} innocuous to neutrino oscillations. Thus, violation of $CP$ invariance in neutrino oscillations is solely produced by the Dirac phase, which determines whether the $CP$ asymmetries $A^{\rm CP}_{kj}=P(\nu_k\to\nu_j)-P(\bar{\nu}_k-\bar{\nu}_j)$ vanish or not.
\\

Now we separate the couplings that will be used in the calculation of the $\gamma\nu\nu$ vertex, to be carried out in the next section. A major motivation for studying left--right models has been that some varieties give rise~\cite{MohSenMajnu} to masses of Majorana neutrinos. Indeed, the necessity of right--handed neutrinos in this formulation is the origin of the celebrated seesaw mechanism~\cite{MohSen}.  The nature of the scalar multiplets of a given left--right model determines~\cite{LgckrSkr} whether neutrino masses are Majorana or Dirac. The Higgs triplets $\Delta_L$ and $\Delta_R$, considered in the present paper, yield neutrino masses of Majorana type.
In a general context, the PMNS matrix mixing left--handed neutrinos may be different to the one that mixes right--handed neutrinos.
For now, we maintain this possibility, so we assume the relations
\begin{eqnarray}
\nu_{j,L}=\sum_{\alpha=1,2,3}{\cal L}_{j\alpha}\,\nu_{\alpha,L}\,,
\\ \nonumber \\
\nu_{j,R}=\sum_{\alpha=1,2,3}{\cal R}_{j,\alpha}\,\nu_{\alpha,L}\,,
\end{eqnarray}
where ${\cal L}$ and ${\cal R}$ denote, respectively, the left and right PMNS matrices~\cite{Branco:1982wp}. The leptonic kinetic terms of the left--right model considered here can be written as
\begin{equation}
i\,\bar{L}_L\gamma^\mu D_\mu L_L+i\,\bar{L}_R\gamma^\mu D_\mu L_R=i\,\bar{l}_kD^\mu\gamma_\mu l_k
+W^+_{a\,\mu}\,\bar{\nu}_\alpha\gamma^\mu\left( v_{a,\alpha k}-a_{a,\alpha k}\gamma^5 \right)l_k+W^-_{a\,\mu}\,\bar{l}_k\gamma^\mu\left( v^*_{a, \alpha k}-a^*_{a,\alpha k}\gamma^5 \right)\nu_\alpha+\cdots,
\label{lrlc}
\end{equation}
where any pair of repeated indices indicates a sum. Here, $a=1,2$, $\alpha=1,2,3$, $j=e,\mu,\tau$, while $D^\mu=\partial^\mu+ie\,A^\mu$ is the U(1)$_e$ covariant derivative. The couplings $v_{a,\alpha k}$ and $a_{a,\alpha k}$ are  given by
\begin{eqnarray}
v_{1,\alpha k}&=&\frac{1}{2\sqrt{2}}\left[ g_R\,e^{i\omega}\sin\zeta\,{\cal R}^*_{k\alpha}+g_L\,\cos\zeta\,{\cal L}^*_{k\alpha} \right],
\label{coupd1}
\\ \nonumber \\
v_{2,\alpha k}&=&\frac{1}{2\sqrt{2}}\left[ g_R\,e^{i\omega}\cos\zeta\,{\cal R}^*_{k\alpha}-g_L\,\sin\zeta\,{\cal L}^*_{k\alpha} \right],
\\ \nonumber \\
a_{1,\alpha k}&=&\frac{1}{2\sqrt{2}}\left[ -g_R\,e^{i\omega}\sin\zeta\,{\cal R}^*_{k\alpha}+g_L\,\cos\zeta\,{\cal L}^*_{k\alpha} \right],
\\ \nonumber \\
a_{2,\alpha k}&=&\frac{1}{2\sqrt{2}}\left[ -g_R\,e^{i\omega}\,\cos\zeta\,{\cal R}^*_{k\alpha}-g_L\,\sin\zeta\,{\cal L}^*_{k\alpha} \right].
\label{coupd4}
\end{eqnarray}
The ellipsis in Eq.~(\ref{lrlc}) represent other terms that include the kinetic terms of neutrinos and couplings involving the neutral massive gauge bosons $Z_\mu$ and $Z'_\mu$. On the other hand, the sum of the SU(2)$_L$ and SU(2)$_R$ pure--gauge terms can be expanded to get
\begin{eqnarray}
-\frac{1}{2}{\rm Tr}\Big[ W_R^{\mu\nu}W_{R\,\mu\nu} \Big]-\frac{1}{2}{\rm Tr}\Big[ W_L^{\mu\nu}W_{L\,\mu\nu} \Big]&=&ie\Big[ \left( W^{+\mu\nu}_aW^-_{a\,\mu}-W^{-\mu\nu}_aW^+_{a\,\mu} \right)A_\nu
+W^+_{a\,\mu}W^-_{a\,\nu}F^{\mu\nu}\Big]+\cdots,
\label{WWAcoup}
\end{eqnarray}
where only the trilinear gauge couplings $WW\gamma$ have been written explicitly and, again, any pair of repeated indices denotes a sum. As it can be appreciated from this expression, there are no mixings of mass eigenstates $W_1$ and $W_2$ in these interactions. Such interesting mixings might be produced, for instance, in effective theories since, due to the mechanism producing the non--manifest left--right symmetry scenario, high--energy scales suppressing nonrenormalizable left and right pure--gauge interactions are, in general, different of each other.

\section{Magnetic moment of neutrinos}
\label{namm}
The determination of whether neutrinos are Majorana or Dirac particles might be attained by experiments that are trying to measure the elusive $2\beta_{0\nu}$ decay. This process consists in the simultaneous $\beta$ decays of two neutrons or two protons, within the same nucleus, which interchange a virtual neutrino, so that the final state comprises the same nucleus and two beta particles, but no neutrinos. For the propagation of such virtual neutrino to happen, the total lepton number must be violated by two units, so that $2\beta_{0\nu}$ decay is forbidden in the context of the SM, which, on the other hand, assumes neutrinos to be massless and thus does not properly describe them. In particular, massive neutrinos of Majorana type gather the sufficient and necessary conditions that allow the occurrence of $2\beta_{0\nu}$ decay. Indeed, the observation of this rare decay would indicate that neutrinos are Majorana particles. Despite experimental efforts, so far no measurement~\cite{KamLandZenno2beta0} of a $2\beta_{0\nu}$ decay has been accomplished, and current lower bounds on the half--life are of order $10^{25}$\,years.
\\

While the observation of the $2\beta_{0\nu}$ decay would be a proof that neutrinos are Majorana--type\footnote{Note that, in left--right, the presence of the $W_R$ and heavy neutrinos might be the cause behind a $2\beta_{0\nu}$ decay measurement~\cite{TNNSV}.}, studies about the $CP$--violating phases are interesting as the cases of Dirac and Majorana neutrinos feature a different number of them. The amplitude of a $2\beta_{0\nu}$ decay is proportional to the effective Majorana mass,
\begin{equation}
m_{2\beta}=\sum_{\alpha=1}^3U^2_{e\alpha}\,m_\alpha,
\end{equation}
which is given in terms of PMNS matrix elements. So, in general, this amplitude depends on the $CP$--violating phases of the PMNS matrix, both of Dirac and Majorana type. The possibility of bounding Majorana phases by means of the $2\beta_{0\nu}$ decay has been considered in Refs.~\cite{DPS,SBFG}. There are other approaches to explore these phases, which naturally appear in physical observables associated to violation of $CP$ invariance. For instance, it has been pointed out that, in the presence of Majorana neutrinos, electric dipole moments of charged leptons depend on these $CP$--violating phases, and that precise measurements of such electric dipoles might be a way to be sensitive to  Majorana phases~\cite{deGG}.
\\

\begin{figure}[!ht]
\center
\includegraphics[width=7cm]{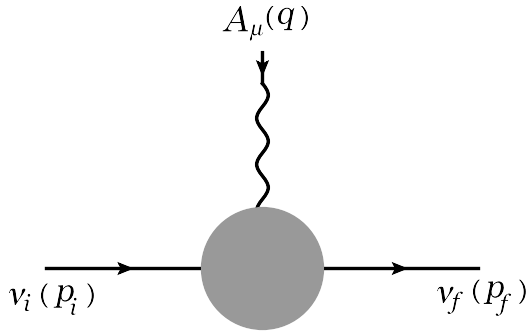}
\caption{\label{nnA} The $\gamma\nu\nu$ vertex.}
\end{figure}

Assuming that neutrinos are massive, that Lorentz and U(1)$_e$ symmetries hold, and following the conventions shown in Fig.~(\ref{nnA}),
it is found\footnote{For general and detailed discussions on this parametrization, see Ref.~\cite{BGS}.} that the general parametrization of the $\gamma\nu\nu$ electromagnetic current has the following structure:
\begin{equation}
\Lambda^{fi}_\mu(q^2)=\left( \gamma_\mu-\frac{q_\mu\slashed{q}}{q^2} \right)\left[ f_Q^{fi}(q^2)+f^{fi}_A(q^2)q^2\gamma^5 \right]+i\,\sigma_{\mu\nu}q^\nu\left[ f^{fi}_M(q^2)+i\,f^{fi}_E(q^2)\gamma^5 \right],
\label{nngvertex}
\end{equation}
where the $i$ and $f$ indices take values corresponding to initial and final neutrino states. The parameters $f_Q^{fi}$, $f_A^{fi}$, $f_M^{fi}$ and $f_E^{fi}$, which depend only on the squared photon momentum $q^2$, are, respectively, the charge, anapole, magnetic dipole and electric dipole form factors. If a real photon is considered, which imposes the condition $q^2=0$, they define the neutrino charge, $q$, anapole moment, $a$, magnetic moment, $\mu$, and electric dipole moment, $\epsilon$:
\begin{equation}
f_Q^{fi}(0)=q_{fi},\hspace{0.5cm}f_A^{fi}(0)=a_{fi},\hspace{0.5cm}f_M^{fi}(0)=\mu_{fi},\hspace{0.5cm}f_E^{fi}(0)=\epsilon_{fi}.
\end{equation}
Those factors for which $i=f$ are known as {\it diagonal electromagnetic form factors}, while those for which $i\ne f$ are called {\it transition electromagnetic form factors}.
\\

The structure of the $\gamma\nu\nu$ electromagnetic vertex, given in Eq.~(\ref{nngvertex}), holds regardless of whether the neutrinos are Dirac or Majorana particles. However, it must be kept in mind that form factors of Dirac neutrinos have different properties than those corresponding to Majorana neutrinos~\cite{empDvsM}. In the case of Majorana neutrinos, the condition $\nu=\nu^{\rm c}$ yields an exact cancellation of all the diagonal form factors, but the anapole, so that Majorana neutrinos can only have transition electromagnetic form factors and the diagonal anapole form factor. Moreover, in the case of transition form factors of Majorana neutrinos, contributions to the $\gamma\nu\nu$ vertex that preserve invariance under $CP$ involve either the electric dipole or the magnetic dipole form factor. If the magnetic dipole form factor is the one that arises in a given calculation, the anapole form factor is forbidden, while a nonzero electric dipole moment solely allows a non--vanishing anapole form factor. Contrastingly, all the diagonal from factors of Dirac neutrinos are, in general, nonzero and all form factors, but the diagonal electric dipole, are allowed if $CP$ is conserved. It is worth emphasizing, from this discussion, that non--diagonal magnetic moments of neutrinos may involve $CP$--violating effects and, thus, they may depend, in general, on the $CP$--violating phases.
\\

Just for a moment, forget about the left--right model and consider some given formulation that involves three massive neutrinos $\nu_\alpha$, with masses $m_\alpha$, and a set of charged gauge bosons $W_{a\,\mu}$, whose masses are denoted by $m_a$. Then assume that these particles gather with the ordinary charged leptons $l_k$, with masses $m_k$, to define charged currents of the form
\begin{equation}
W^+_{a\,\mu}\,\bar{\nu}_\alpha\gamma^\mu\left( v_{a,\alpha k}-a_{a,\alpha k}\gamma^5 \right)l_k+W^-_{a\,\mu}\,\bar{l}_k\gamma^\mu\left( v^*_{a, \alpha k}-a^*_{a,\alpha k}\gamma^5 \right)\nu_\alpha,
\label{generalc}
\end{equation}
where pairs of repeated indices always indicate sums. Here, $\alpha=1,2,3$, $j=e,\mu,\tau$, while the index $a$ labels all the charged gauge bosons, from the SM and from some new--physics description as well. Note that the charged currents given in Ec.~(\ref{lrlc}), which correspond to the left--right model, fit Ec.~(\ref{generalc}). Also assume a general $\gamma WW$ coupling, like that of Eq.~(\ref{WWAcoup}).
\begin{figure}[!ht]
\center
\includegraphics[width=7cm]{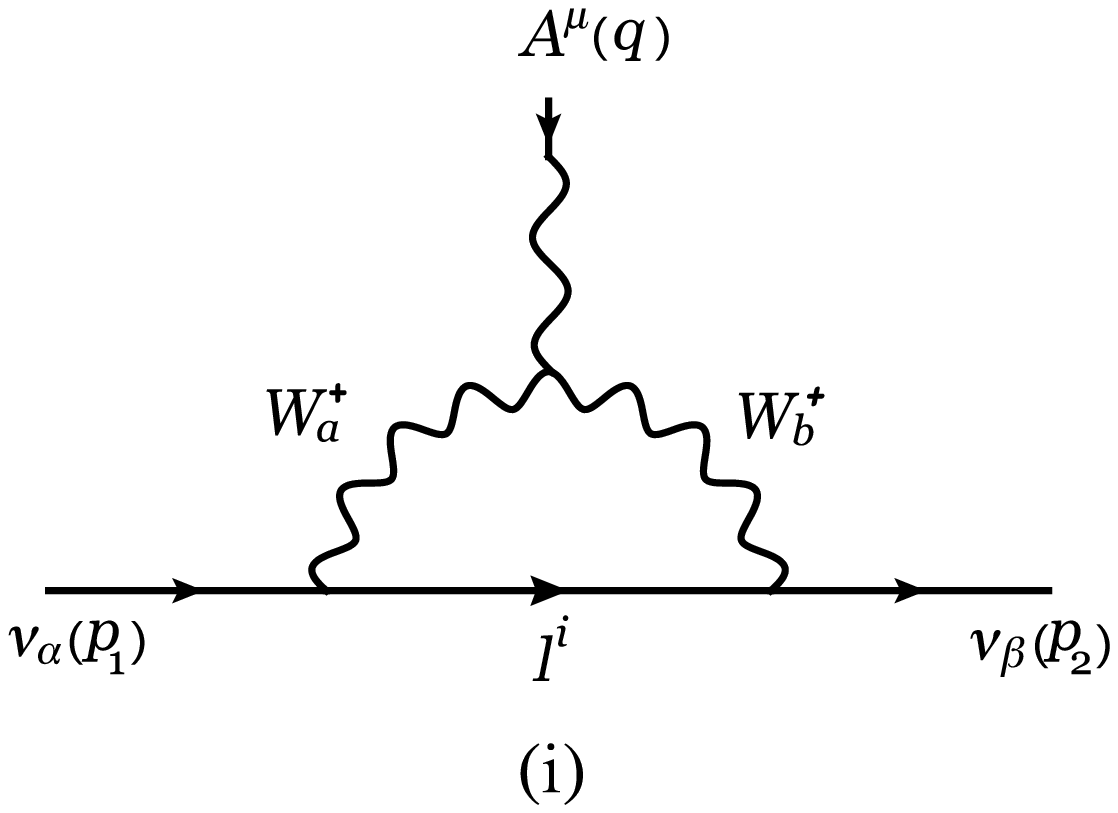}
\hspace{1cm}
\includegraphics[width=7cm]{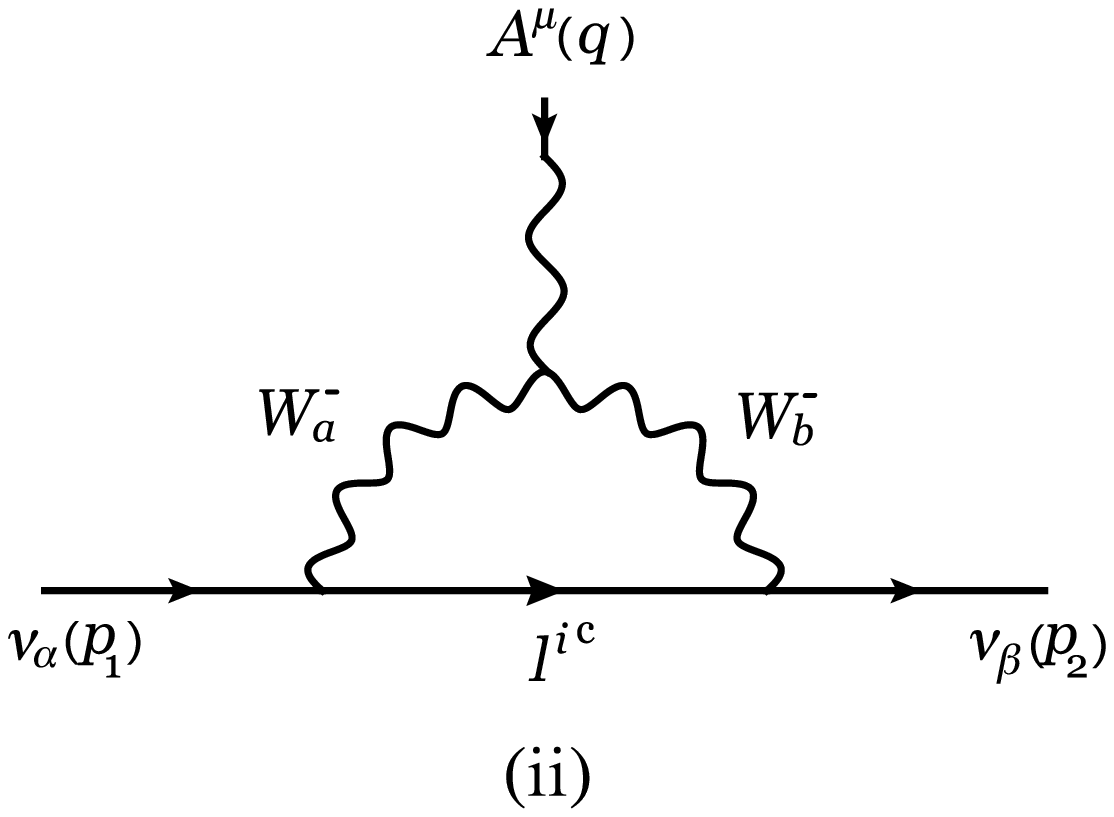}
\\
\vspace{0.7cm}
\includegraphics[width=7cm]{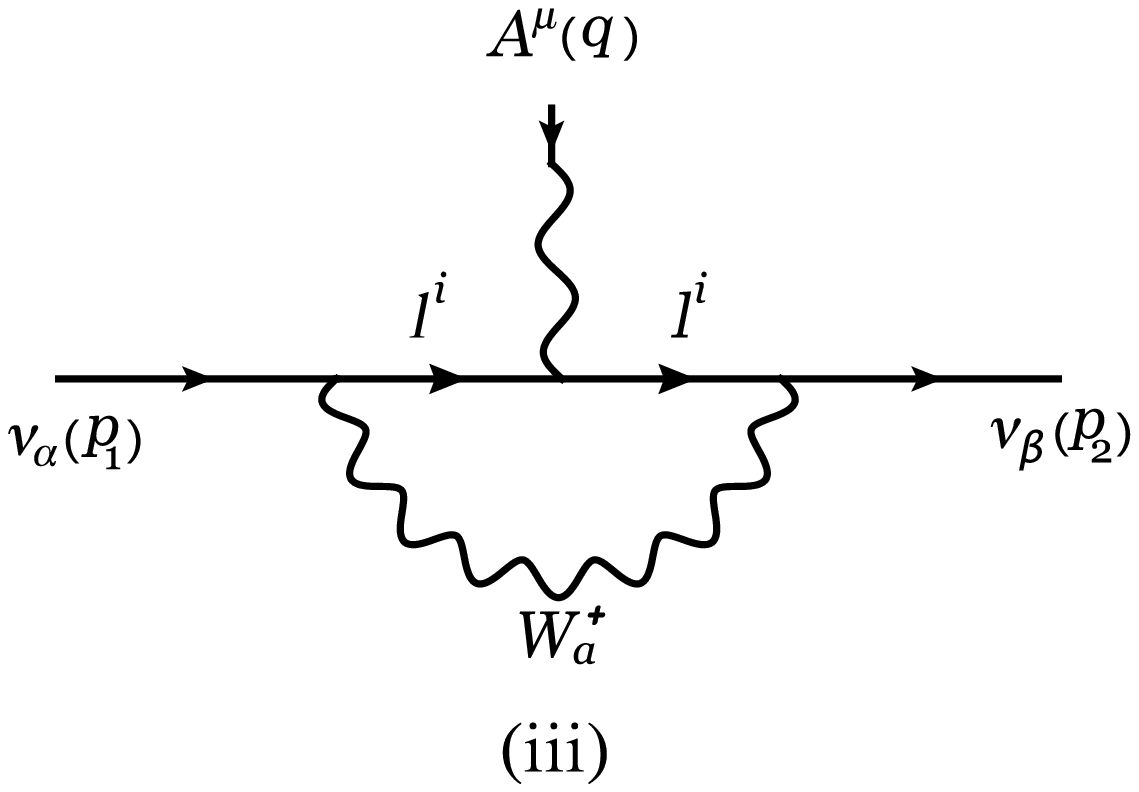}
\hspace{1cm}
\includegraphics[width=7cm]{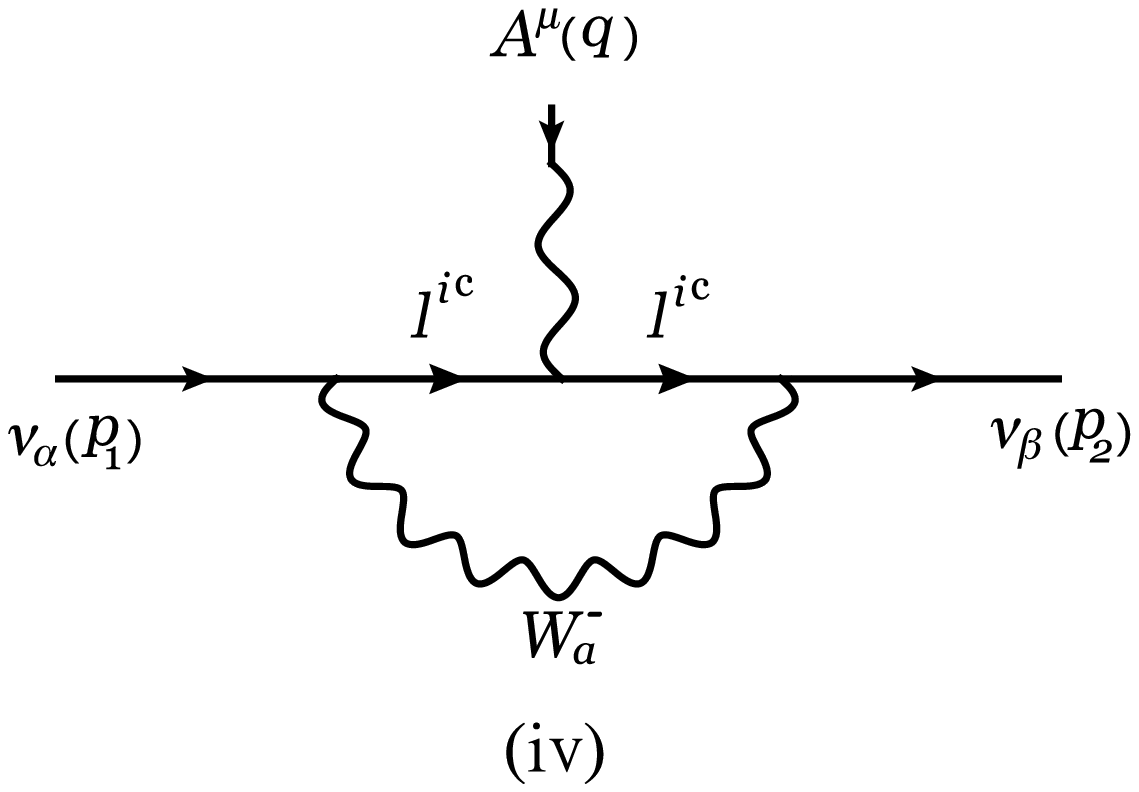}
\caption{\label{contriagrams} One--loop diagrams with internal charged gauge bosons.}
\end{figure}
The one--loop contributions from these general couplings to the vertex $\gamma\nu\nu$ are given by the Feynman diagrams of Fig.~(\ref{contriagrams}).
As we commented before, neutrinos in the left--right model with scalar Higgs triplets are described by Majorana fields. The Majorana condition, $\nu=\nu^{\rm c}$, gives rise to important differences, with respect to the case of Dirac neutrinos, of the manner~\cite{Bkay} in which contributions to the neutrino electromagnetic vertex must be calculated. While diagrams (i) and (iii) of Fig.~(\ref{contriagrams}), corresponding to external neutrinos, should be calculated if we were dealing with Dirac neutrinos, diagrams (ii) and (iv), which involve external antineutrinos, should not. Nevertheless, for Majorana neutrinos the consideration of all four diagrams is mandatory. The result of performing a $CPT$ transformation on Eq.~(\ref{generalc}) is
\begin{eqnarray}
(CPT)^{-1}\Big[ W^+_{a\,\mu}\,\bar{\nu}_\alpha\gamma^\mu(v_{a,\alpha k}-a_{a,\alpha k}\gamma^5)l_k
\nonumber \\ \nonumber \\
+W^-_{a\,\mu}\,\bar{l}_k\gamma^\mu(v^*_{a,\alpha k}-a^*_{a,\alpha k}\gamma^5)\nu_\alpha \Big]CPT&=&-W^+_{a\,\mu}\,\bar{l}^{\rm c}_k\gamma^\mu(v_{a,\alpha k}+a_{a,\alpha k}\gamma^5)\nu^{\rm c}_\alpha
\nonumber \\ \nonumber \\ &&
-W^-_{a\,\mu}\,\bar{\nu}^{\rm c}_\alpha\gamma^\mu(v^*_{a,\alpha k}+a^*_{a,\alpha k}\gamma^5)l^{\rm c}_k.
\label{lrlcCPT}
\end{eqnarray}
By virtue of the Majorana condition, $\nu^{\rm c}=\nu$, these couplings generate contributions to the $\gamma\nu\nu$ vertex, for they become building blocks of diagrams (ii) and (iv), in Fig.~(\ref{contriagrams}). Now consider the transformation under $CP$ of Eq.~(\ref{generalc}), which yields
\begin{eqnarray}
(CP)^{-1}\Big[ W^+_{a\,\mu}\,\bar{\nu}_\alpha\gamma^\mu\left( v_{a,\alpha k}-a_{a,\alpha k}\gamma^5 \right)l_k
\nonumber \\ \nonumber \\
+W^-_{a\,\mu}\,\bar{l}_k\gamma^\mu\left( v^*_{a, \alpha k}-a^*_{a,\alpha k}\gamma^5 \right)\nu_\alpha \Big]CP&=&W^+_{a\,\mu}\,\bar{\nu}_\alpha\gamma^\mu\left( v^*_{a,\alpha k}-a^*_{a,\alpha k}\gamma^5 \right)l_k
\nonumber \\ \nonumber \\ &&
+W^-_{a\,\mu}\,\bar{l}_k\gamma^\mu\left( v_{a,\alpha k}-a_{a,\alpha k}\gamma^5 \right)\nu_\alpha.
\label{CPtrans}
\end{eqnarray}
As it can be appreciated from this expression, tree--level $CP$ violation arises from the general charged currents of Eq.~(\ref{generalc}) as long as the couplings $v_{a,\alpha k}$ and $a_{a,\alpha k}$ are complex quantities.% but $CP$ is conserved if they are real.
\\

Using Eqs.~(\ref{WWAcoup}), (\ref{generalc}) and (\ref{lrlcCPT}), we write the the total contributions from diagrams of the form (i) or (ii) generically as
\begin{eqnarray}
\Lambda^{\beta\alpha}_\mu(q^2)&=&(-1)^z\,ie\sum_{a,b}\sum_i\delta^{ab}\mu^{4-D}\int\frac{d^Dk}{(2\pi)^D}\frac{\bar{u}_\beta(p_2)\gamma_\sigma(V_{b,\beta i}+A_{b,\beta i}\,\gamma^5)(\slashed{k}+m_i)\gamma_\lambda(V^*_{a,\alpha i}+A^*_{a,\alpha i}\,\gamma^5)u_\alpha(p_1)}{\Big[ (k-p_1)^2-m_a^2 \Big]\Big[ (k-p_2)^2-m_b^2 \Big]\Big[ k^2-m_i^2 \Big]}
\nonumber \\ \nonumber \\ && \times
\bigg[ g^{\sigma\rho}-(1-\xi)\frac{(k-p_2)^\sigma(k-p_2)^\rho}{(k-p_2)^2-\xi m_b^2} \bigg]\bigg[ g^{\nu\lambda}-(1-\xi)\frac{(k-p_1)^\nu(k-p_1)^\lambda}{(k-p_1)^2-\xi m_a^2} \bigg]
\nonumber \\ \nonumber \\ && \times
\Big[ (2p_2-p_1-k)_\nu\,g_{\mu\rho}+(2p_1-p_2-k)_\rho\,g_{\nu\mu}+(2k-p_1-p_2)_\mu\,g_{\rho\nu} \Big],
\label{1stvf}
\end{eqnarray}
where $\xi$ is the gauge--fixing parameter. The loop integral is carried out within the dimensional regularization approach, so we have set it in an arbitrary dimension, $D$. The factor $\mu^{4-D}$ is then intended to appropriately correct units of the $D$--dimensional loop integral. Note that, in the right--hand side of Eq~(\ref{1stvf}), $z=1$ for diagrams (i) and $z=0$ for diagrams (ii). The definitions of the factors $V_{a,\alpha i}$ and $A_{a,\alpha i}$ are given in Table~\ref{defs}. The indices $\alpha=1,2,3$ and $\beta=1,2,3$ are used to label different neutrino mass eigenstates. All other greek indices denote space--time components. The index $i$ corresponds to different charged leptons, and the indices $a=1,2$ and $b=1,2$ label mass eigenstates of charged gauge bosons, $W_1$ and $W_2$. The generic loop integral appearing in this vertex function has terms with four and five poles, but it can be decomposed into a sum of four terms, each one involving just three poles.
\begin{table}[ht]
\centering
\begin{tabular}{| c | c | c | c|}
\hline
& $V_{a,\alpha i}$ & $A_{a,\alpha i}$ & $(-1)^z$
\\ \hline
Diagrams (i) and (iii) & $v_{a,\alpha i}$ & $-a_{a,\alpha i}$ & $-1$
\\[0.2cm]
Diagrams (ii) and (iv) & $v^*_{a,\alpha i}$ & $a^*_{a,\alpha i}$ & $+1$
\\ \hline
\end{tabular}
\caption{\label{defs} Couplings and global factors of loop integrals in terms of the couplings defined in Eqs.~(\ref{coupd1}) to (\ref{coupd4}).}
\end{table}
Similarly, the total contributions from diagrams of the form (iii) or (iv) can be expressed as
\begin{eqnarray}
\bar{\Lambda}^{\beta\alpha}_\mu(q^2)&=&-(-1)^z\,ie\sum_a\sum_i\mu^{4-D}\int\frac{d^Dk}{(2\pi)^D}\bigg[ g^{\rho\nu}-(1-\xi)\frac{k^\rho k^\nu}{k^2-\xi m_a^2} \bigg]
\nonumber \\ \nonumber \\ && \times
\frac{\bar{u}_\beta(p_2)\gamma_\rho(V_{a,\beta i}+A_{a,\beta i}\gamma^5)(\slashed{k}-\slashed{p}_2-m_i)\gamma_\mu(\slashed{k}-\slashed{p}_1-m_i)\gamma_\nu(V^*_{a,\alpha i}+A^*_{a,\alpha i}\gamma^5)u_\alpha(p_1)}{\Big[ (k-p_1)^2-m_i^2 \Big]\Big[ (k-p_2)^2-m_i^2 \Big]\Big[ k^2-m_a^2 \Big]},
\end{eqnarray}
where we have employed the notation of the vertex functions for diagrams (i) and (iii), given in Eq.~(\ref{1stvf}). In this case, however, $z=1$ for diagrams (iii) and $z=0$ for diagrams (iv).
\\

To solve the loop integrals we have assumed that the photon is off shell, so that $q^2\ne0$. We have followed the Feynman parameters technique and, using dimensional regularization, we have isolated the ultraviolet--diverging terms, which exactly vanish in the magnetic dipole and electric dipole form factors, so that the total, and finite, contributions from diagrams of Fig.~(\ref{contriagrams}) to these form factors have the following structure:
\begin{eqnarray}
f^{\beta\alpha}_M(q^2)&=&\frac{m_e}{8\pi^2}\,\mu_{\rm B}\sum_i\sum_a\Big[ \left( a_{a,\beta i}\;a^*_{a,\alpha i}-a^*_{a,\beta i}\;a_{a,\alpha i} \right)I_{1}+\left( v_{a,\beta i}\;v^*_{a,\alpha i}-v^*_{a,\beta i}\;v_{a,\alpha i} \right)I_{2} \Big],
\label{mdff}
\\ \nonumber \\
f_E^{\beta\alpha}(q^2)&=&\frac{e}{(4\pi)^2}\sum_i\sum_a\Big[ \left( a_{a,\beta i}\;v^*_{a,\alpha i}+a^*_{a,\beta i}\;v_{a,\alpha i} \right)I_3+\left( v_{a,\beta i}\;a^*_{a,\alpha i}+v^*_{a,\beta i}\;a_{a,\alpha i} \right)I_4 \Big],
\label{edff}
\end{eqnarray}
where $\mu_{\rm B}$ is the Bohr magneton and  $I_1$, $I_2$, $I_3$ and $I_4$ are parametric integrals which depend on the gauge--fixing parameter $\xi$.
 %No particular gauge was chosen to carry out the calculations, so that gauge dependence appears in the parametric integrals through the gauge--fixing parameter $\xi$. In this general context, the integrands of the parametric integrals are enormous and complicated functions of the squared photon momentum $q^2$, of the masses of neutrinos, charged leptons and gauge bosons, and, of course, of the gauge--fixing parameter. For this reason, we do not provide the explicit form of such integrals. As it occurs in the case of $f_M^{\beta\alpha}$ and $f_E^{\beta\alpha}$, the contributions to the charge and anapole form factors are given in terms of complicated parametric integrals.
Despite the intricate structure of all integrals, one can verify that some properties~\cite{BGS} of the electromagnetic form factors hold. From the factors multiplying parametric integrals in Eq.~(\ref{mdff}), it is explicit that diagonal magnetic dipole form factors, for which $\beta=\alpha$, vanish. A cancellation of diagonal electric dipole form factors also happens, but it cannot be determined simply from the general structure of Eq.~(\ref{edff}), since it requires manipulations of the integrands of $I_3$ and $I_4$. We have verified that the contributions to diagonal charge form factors vanish as well, but contributions to diagonal anapole form factors remain. On the other hand, notice that the only sources of $CP$--violation in diagrams of Fig.~(\ref{contriagrams}) are the couplings $Wl\nu$. As we explicitly showed in Eq.~(\ref{CPtrans}), violation of $CP$ invariance in such couplings requires the coefficients $v_{a,\alpha i}$ and $a_{a,\alpha i}$ to be complex quantities. Otherwise, $CP$ is conserved in these interactions and, consequently, in the corresponding contributions to the $\gamma\nu\nu$ electromagnetic vertex. Keeping this in mind, it can be appreciated, from the general structure of Eq.~(\ref{mdff}), that the assumption of $CP$ invariance yields a cancellation of the magnetic dipole form factor. Moreover, we have verified that $CP$--conservation consistently yields an analogous elimination of contributions to charge form factors, but electric dipole and anapole form factors perdure.
\\

%The magnetic dipole form factor $f^{\beta\alpha}_M$, given in Eq.~(\ref{mdff}), can be split into the sum of contributions from diagrams with external neutrinos and those from diagrams with external antineutrinos, which is inherited by the corresponding magnetic moment $\mu_{\beta\alpha}$.

The contributions to the neutrino electromagnetic vertex that originate in the left--right model considered in the present study enter through one--loop diagrams like those of Fig.~(\ref{contriagrams}), which in this particular context incorporate only the charged $W_1$ and $W_2$ bosons. Additionally, diagrams involving neutral $Z$ and $Z'$ bosons, physical scalar fields, and pseudo--Goldstone bosons contribute. Up to this point, we have performed a general calculation of the contributions from charged gauge bosons in the $R_\xi$ gauge, but in the end we will get rid of spurious degrees of freedom by taking the unitary gauge, so that a calculation of diagrams involving pseudo--Goldstone bosons is not necessary. Furthermore, since we are particularly interested in the contributions to the magnetic dipole form factor, we neglect those diagrams involving $Z$ and $Z'$ bosons. Finally, it has been claimed that diagrams with physical scalars produce contributions to the magnetic dipole form factor that are small in comparison with the ones coming from diagrams with charged gauge bosons\footnote{This is to be contrasted with Ref.~\cite{BB}, where a calculation of the neutrino MMs in left--right was performed with special emphasis on the light--heavy neutrino mixing. The authors report that contributions from singly--charged scalars may be even larger than those from charged gauge bosons. Our analytic expressions for the $W$ bosons contributions coincide, in the unitary gauge, with those given in this reference.}~\cite{CGZ}. Thus, we consider exclusively the contributions from the diagrams shown in Fig.~(\ref{contriagrams}). In the case of the electric dipole moment, a GIM cancellation takes place, which severly attenuates the contributions to this electromagnetic factor. In view of this, we will not further analyze the electric dipole moment.
 \\

In the next step, we insert the definitions of the coefficients $v_{a,\alpha k}$ and $a_{a,\alpha k}$ for the left--right model, which are listed in Eqs.~(\ref{coupd1}) to (\ref{coupd4}), into Eq.~(\ref{mdff}) and set $q^2=0$, in order to obtain the expression for the neutrino magnetic moments.
 %One of the main motivations to explore contributions from different SM extensions to neutrino MMs is that the sole addition of massive neutrinos to the SM generates~\cite{BGS} contributions that are strongly suppressed by tiny neutrino masses, which yields a huge difference between theoretical predictions and experimental sensitivity. The left--right model, contrastingly, gives rise to neutrino magnetic moments incorporating terms that are independent of such neutrino masses and thus are much larger. These dominant contributions come from diagrams (i) and (ii) with $a=b=1$. Taking into account that the masses of the charged gauge bosons are much larger than the masses of light neutrinos and charged leptons,
 We solve parametric integrals and find that the dominant contributions can be written as
\begin{equation}
\mu_{\beta\alpha}^{\rm M}\approx i\,\mu_{\rm B} \frac{g_L g_R}{(4\pi)^2}\sin\zeta \cos\zeta\, m_e\frac{m_2^2-m_1^2}{m_2^2m_1^2}\frac{2\xi^2-3\xi-\xi\log\xi+1}{(\xi-1)^2}\sum_{i=e,\mu,\tau}m_i\,{\rm Im}\Big[ e^{i\omega}\left( {\cal R}^\dag_{\alpha i}\,{\cal L}_{i\beta}-{\cal R}^\dag_{\beta i}\,{\cal L}_{i\alpha} \right) \Big].
\label{mdm1}
\end{equation}
From this expression, it is clear that the resulting MM is an imaginary quantity, which is consistent with the general properties of the electromagnetic form factors of Majorana neutrinos~\cite{BGS}. Notice that $\mu^{\rm M}_{\beta\alpha}=-\mu^{\rm M}_{\alpha\beta}$, which means that there are only three independent neutrino MMs.
\\

The mixings of Majorana left--handed neutrinos can be parametrized by a set of three mixing angles, one Dirac phase and two Majorana phases. In general, there is an analogous, but independent, set for right--handed neutrinos. The left and right Majorana mixing matrices can be expressed as ${\cal L}=V_L\,P_L$ and ${\cal R}=V_R\,P_R$. The Majorana phases $l_2$, $l_3$, $r_2$ and $r_3$ are incorporated by the matrices $P_L={\rm diag}(1,e^{i\frac{l_2}{2}},e^{i\frac{l_3}{2}})$ and $P_R={\rm diag}(1,e^{i\frac{r_2}{2}},e^{i\frac{r_3}{2}})$, while the rest of the mixing parameters are located in the matrices $V_L$ and $V_R$. From Eq.~(\ref{mdm1}), notice that the  ${\cal L}$ and ${\cal R}$ mixing matrices enter the MMs $\mu_{\beta\alpha}^{\rm M}$ only in products of two of them, so that the result can be written in terms of Majorana phase differences, $\phi_{\beta\alpha}=(l_\beta-r_\alpha)/2$, with $l_1=0$ and $r_1=0$.
Recall that the contributions that generate the MMs exhibited in Eq.~(\ref{mdm1}) come from diagrams with external neutrinos and from diagrams involving external antineutrinos as well. If neutrinos in this formulation were Dirac--like, only the former type of diagrams should be calculated, and the corresponding Dirac MMs, which we denote by $\mu_{\beta\alpha}^{\rm D}$, would be
\begin{equation}
\mu^{\rm D}_{\beta\alpha}\approx\mu_{\rm B}\frac{g_L g_R}{2(4\pi)^2}\sin\zeta\,\cos\zeta\, m_e\frac{m_1^2-m_2^2}{m_1^2m_2^2}\frac{2\xi^2-3\xi-\xi\log\xi+1}{(\xi-1)^2}\sum_{i=e,\mu,\tau}m_i\Big[ e^{i\omega}\,{\cal R}^\dag_{\beta i}\,{\cal L}_{i\alpha}+e^{-i\omega}\,{\cal L}^\dag_{\beta i}\,{\cal R}_{i\alpha} \Big].
\label{DiracMM}
\end{equation}
Clearly, the $\mu^{\rm D}_{\beta\alpha}$ contributions are complex quantities, contrastingly to the case of the MMs $\mu_{\beta\alpha}^{\rm M}$, which are purely imaginary. Notice that the Dirac case forbids the presence of Majorana $CP$ phases. If we take the Majorana phases equal to zero in $\mu^{\rm M}_{\beta\alpha}$, we get te simple relation
\begin{equation}
\left. \mu^{\rm M}_{\beta\alpha}\right|_{\phi_{\beta\alpha}=0}=\mu^{\rm D}_{\beta\alpha}-(\mu^{\rm D}_{\beta\alpha})^*.
\end{equation}
This sum consistently eliminates the real parts of the MMs and renders the resulting expression imaginary.

\section{Estimations and discussion of results}
\label{discussion}
In this section we discuss our results. To this aim, we explore two scenarios, distinguished of each other by different shapes of the right PMNS matrix, ${\cal R}$. The first case that we take into account is a particular sort of maximal mixing, defined by the assumption that the left and right mixings are equal. For the second ${\cal R}$ shape, we consider a CKM--like right mixing, which we assume to be close to the identity matrix, except for the $CP$ phases, which remain general.

\subsection{Maximal right mixing}
Here we assume that left-- and right--handed neutrinos share the same mixing matrix, that is, ${\cal L}={\cal R}$. In such case, the Majorana phases for left-- and right--handed neutrino states are the same, and, in what follows, we denote them by $\varphi_\alpha=l_\alpha=r_\alpha$. Furthermore, any Majorana phase difference $\phi_{\beta\alpha}$ is antisymmetric with respect to $\beta$ and $\alpha$ and vanishes for $\beta=\alpha$. Now we eliminate the unphysical degrees of freedom by taking the unitary gauge, $\xi\to\infty$, and express the neutrino MM as
\begin{equation}
\mu_{\beta\alpha}^{\rm M}=i\,\mu_{\rm B}\frac{g_Lg_R}{(2\pi)^2}\sin\zeta\cos\zeta\,m_e\frac{m_2^2-m_1^2}{m_2^2m_1^2}\cos\omega\Big[ \sin\phi_{\beta\alpha}\,m_{1,{\beta\alpha}}+\sin(\delta+\phi_{\beta\alpha})\,m_{2,\beta\alpha}-\sin(\delta+\phi_{\alpha\beta})\,m_{2,\alpha\beta} \Big],
\label{mdm2}
\end{equation}
where the coefficients $m_{j,\beta\alpha}$ are defined in terms of masses of charged leptons and mixing angles $\theta_{12}$, $\theta_{23}$ and $\theta_{13}$. Their explicit expressions are provided in Appendix \ref{mcoefs}.
The MM that we showed in Eq.~(\ref{mdm2}) is given explicitly in terms of the Dirac $CP$--violating phase, $\delta$, and phase differences, $\phi_{\beta\alpha}$, of Majorana phases $\varphi_\alpha$. It also depends on the complex phase $\omega$, which originates in the second stage of symmetry breaking and is incorporated to the charged currents as a consequence of the mixing of left and right charged gauge bosons.
For $\omega=\pm\pi/2$, these leading contributions cancel, despite $CP$ violation is present. This feature is not general, for it arises as a consequence of our assumption of equal left and right mixings. On the other hand, we are particularly interested in the sensitivity of the neutrino MMs to the PMNS $CP$ phases. For these reasons we assume, from now on, that $\cos \omega=1$, which leaves all $CP$--violation in the hands of the PMNS phases.
%On the other hand, the Dirac--like MMs are given by
%\begin{equation}
%\mu^{\rm D}_{\beta\alpha}=\mu_{\rm B}\frac{g_Lg_R}{8\pi^2}\sin\zeta\cos\zeta\,m_e\frac{m_1^2-m_2^2}{m_2^2m_1^2}\cos\omega\Big[ m_{1,{\beta\alpha}}+e^{-i\delta}\,m_{2,\beta\alpha}+e^{i\delta}\,m_{2,\alpha\beta}  \Big].
%\label{mdm3}
%\end{equation}
\\

The dependence of the $\mu^{\rm M}_{\beta\alpha}$ on the mass of the heavy charged gauge boson improves the contributions to the MMs for larger values of such mass, that is, the larger the $W_2$ mass, the greater the MMs. Small values of the $m_2$ mass, which are already discarded~\cite{OnessEbel}, would play an important role, since in that light--mass region the MMs are very sensitive to them. However, as larger values of the $m_2$ mass are considered, the contributions to MMs soon stabilize and grow very slowly.
A hint pointing towards a TeV--scale mass for the right charged gauge boson was presented recently by the CMS Collaboration~\cite{CMSWrsrch}, which set a lower bound of 3.0\,TeV on this mass.
Moreover, an excess with significance $2.8\sigma$ in two leptons and two jets events that was reported in that paper has been interpreted~\cite{Dparity,WrGluza1,WrGluza2,DM,CDL,YKS,AEM} as a signal of right--handed charged gauge bosons with masses within the range 1.9\,TeV--2.4\,TeV.
\\

Left--right mixing of charged gauge bosons is important to the $\mu_{\beta\alpha}^{\rm M}$, for a nonzero value of the mixing angle $\zeta$ avoids GIM suppression, thus enlarging the contributions to MMs. But whether MMs of neutrinos are actually generated or not is defined by the $CP$--violating phases of the PMNS matrix.
According to the bound~\cite{LgckrSkr} on the mixing angle $\zeta$, to the current best directly measured values~\cite{PartDatGru} of the PMNS mixing angles, and assuming that the $W_2$ mass is within the range of a few TeVs, we write the magnitude of the MM of neutrinos as
\begin{equation}
|\mu^{\rm M}_{\beta\alpha}|<\mu_{\rm B}\,(\,4\times10^{-11}\,{\rm GeV}^{-1})\,\left|\, m_{1,\beta\alpha}\,\sin\phi_{\beta\alpha}+(m_{2,\beta\alpha}-m_{2,\alpha\beta})\sin\delta\cos\phi_{\beta\alpha}+(m_{2,\beta\alpha}+m_{2,\alpha\beta})\cos\delta\sin\phi_{\beta\alpha} \right|.
\label{majMM1stestim}
\end{equation}
The $m_{j,\beta\alpha}$ are of order $10^{-1}$, so that neutrino MMs are, at most, of order $10^{-11}$.
As we pointed out before, if violation of $CP$ invariance were absent, all contributions to the neutrino MM would be eliminated, which evidently occurs with the leading effects shown in Ec.~(\ref{majMM1stestim}). In case that $CP$ invariance is violated, the Dirac and Majorana phases determine the size of the contributions. Up to now, the Dirac phase has not been measured, although studies aimed to find the value of this quantity are available. The best fit $\delta\simeq3\pi/2$ for the Dirac phase was reported in Ref.~\cite{GaMaSc}, where the value $\pi/2$ was disfavored. In Ref.~\cite{MPRS}, on the other hand, the Dirac phase was found to be around $\delta\simeq\pi$ for a bimaximal shape of the PMNS matrix, while for a tribimaximal neutrino mixing this phase is reported to be around $\delta\simeq3\pi/2$ or $\pi/2$, with the exclusion of the values $\delta=0,\pi,2\pi$ at greater than $4\sigma$. The T2K Collaboration recently reported~\cite{T2Kdiracphase} electron--neutrino appearance from a muon--neutrino beam beyond $5\sigma$, finding that the value $\delta=-\pi/2$ is preferred by combined T2K data and reactor measurements.
\\

\begin{figure}[!ht]
\center
\includegraphics[width=13cm]{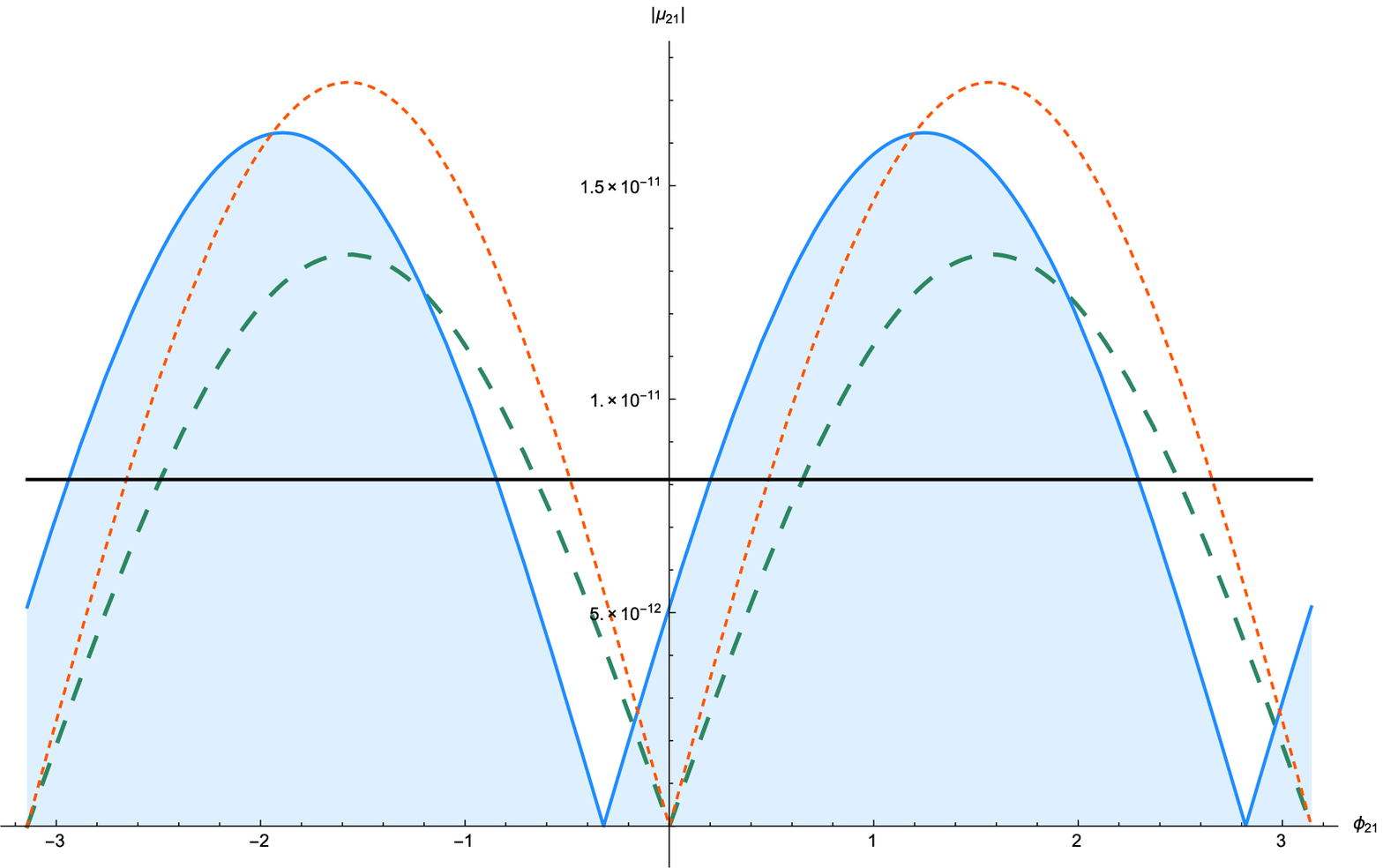}
\hspace{0.5cm}
\includegraphics[width=3cm]{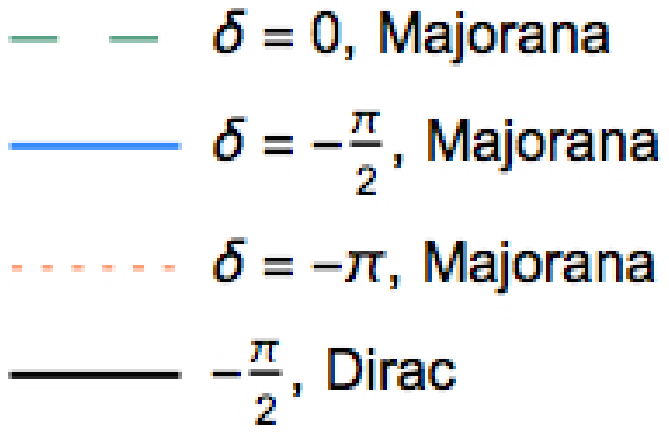}
\caption{\label{phi21} Dependence of the transition Magnetic moment $|\mu^{\rm M}_{21}|$ on the Majorana phase $\varphi_{2}$ for different Dirac phases $\delta$. The upper bound on the Dirac MM magnitude $|\mu_{21}^{\rm D}|$, for $\delta=-\pi/2$, is represented by a horizontal line.}
\end{figure}

For illustrative purposes, we assume that $\delta=-\pi/2$, so that
\begin{equation}
|\mu^{\rm M}_{\beta\alpha}|<\mu_{\rm B}\,(\,4\times10^{-11}\,{\rm GeV}^{-1})\,\left|\, m_{1,\beta\alpha}\,\sin\phi_{\beta\alpha}-(m_{2,\beta\alpha}-m_{2,\alpha\beta})\cos\phi_{\beta\alpha} \right|.
\end{equation}
The behavior of the MMs is exemplified in Fig.~(\ref{phi21}),
which exhibits what happens in case of the magnitude $|\mu_{21}^{\rm M}|$ for different elections of the Dirac phase. The curves represent upper limits on the $|\mu^{\rm M}_{21}|$, for different Majorana phase differences, while the region under these curves (a shaded region in the case of $\delta=-\pi/2$) comprehends all allowed values of the MMs, which can be defined, for instance, by the amount of mixing of left and right charged gauge bosons.
As it can be appreciated from these plots, the value $\delta=-\pi/2$, favored by Ref.~\cite{T2Kdiracphase}, does not produce the largest contributions to the neutrino MMs, since angles such as $-\pi$ yield larger results.
The upper bounds on the MMs reach their maxima at phase differences $\phi^{\rm max}_{\beta\alpha}$ given by
\begin{eqnarray}
\tan\phi^{\rm max}_{\beta\alpha}=&\displaystyle\frac{m_{1,\beta\alpha}}{m_{2,\alpha\beta}-m_{2,\beta\alpha}}, &\displaystyle \,{\rm for} \,\,\,\,\phi_{\beta\alpha}\ne\pm\frac{\pi}{2},
\\ \nonumber \\
\cot\phi^{\rm max}_{\beta\alpha}=&\displaystyle\frac{m_{2,\alpha\beta}-m_{2,\beta\alpha}}{m_{1,\beta\alpha}},&\displaystyle \,{\rm for} \,\,\,\,\phi_{\beta\alpha}\ne\pi,0.
\end{eqnarray}
so that
\begin{equation}
|\mu^{\rm M}_{\beta\alpha}|<\mu_{\rm B}\,(\,4\times10^{-11}\,{\rm GeV}^{-1})\sqrt{m_{\rm 1,\beta\alpha}^2+(m_{2,\alpha\beta}-m_{2,\beta\alpha})^2}.
\end{equation}
Recall that $\varphi_1=0$, so that $\phi_{21}=\varphi_2/2$ and $\phi_{31}=\varphi_3/2$, which means that the MMs $|\mu^{\rm M}_{21}|$ and $|\mu_{32}^{\rm M}|$ grant us direct access to the Majorana phases. From the last equations, we estimate, for $\delta=-\pi/2$, the upper bound on each MM, the $\phi_{\beta\alpha}^{\rm max}$ phase differences yielding these maxima, and the corresponding Majorana phases:
\begin{eqnarray}
|\mu^{\rm M}_{21}|\lesssim1.62\times10^{-11}, & & %\frac{\varphi^{\rm max}_2}{2}=
\phi^{\rm max}_{21}\approx71.62^{\rm o}, -108.29^{\rm o},\,\,\, \varphi_2\approx143.24^{\rm o},
\label{muest1}
\\ \nonumber \\
|\mu^{\rm M}_{31}|\lesssim1.88\times10^{-11},& & %\frac{\varphi^{\rm max}_3}{2}=
\phi^{\rm max}_{31}\approx73.91^{\rm o}, -106^{\rm o},\,\,\,\varphi_3\approx147.82^{\rm o},
\label{muest2}
\\ \nonumber \\
|\mu^{\rm M}_{32}|\lesssim2.76\times10^{-11},& & \phi^{\rm max}_{32}\approx97.4^{\rm o}, -83.08^{\rm o}.
\label{muest3}
\end{eqnarray}
Fig.~(\ref{phi21}) also provides a comparison between the Majorana and Dirac cases. It shows the upper bound on the $|\mu^{\rm D}_{21}|$ magnitude of the Dirac MM, which is obtained by setting the condition ${\cal L}={\cal R}$ in Eq.~(\ref{DiracMM}), for a Dirac phase $\delta=-\pi/2$. Such upper bound is, in this case, a horizontal line defined by $|\mu^{\rm D}_{21}|\approx8.13\times10^{-12}\mu_{\rm B}$.
\\

Is there any set of values for the PMNS phases that violates $CP$ invariance and eliminates all neutrino MMs at the same time?
As Fig.~(\ref{phi21}) illustrates, even in the presence of Dirac $CP$--violation, certain Majorana phase differences may eliminate a particular neutrino MM.
Indeed, at the saddle points, $\phi_{\beta\alpha}^0$, determined by
\begin{eqnarray}
\tan\phi^0_{\beta\alpha}=&-\displaystyle\frac{m_{2,\alpha\beta}-m_{2,\beta\alpha}}{m_{1,\beta\alpha}}, &\displaystyle \,{\rm for} \,\,\,\,\phi_{\beta\alpha}\ne\pm\frac{\pi}{2},
\\ \nonumber \\
\cot\phi^0_{\beta\alpha}=&-\displaystyle\frac{m_{1,\beta\alpha}}{m_{2,\alpha\beta}-m_{2,\beta\alpha}},&\displaystyle \,{\rm for} \,\,\,\,\phi_{\beta\alpha}\ne\pi,0,
\end{eqnarray}
the MM $\mu_{\beta\alpha}^{\rm M}$ vanishes. The corresponding angles are
\begin{eqnarray}
|\mu^{\rm M}_{21}|\approx0, &&\,\,\,\varphi^0_{2}\approx-36.86^{\rm o},
\label{ang21}
\\ \nonumber \\
|\mu^{\rm M}_{31}|\approx0, &&\,\,\,\varphi^0_{3}\approx-31.98^{\rm o},
\label{ang31}
\\ \nonumber \\
|\mu^{\rm M}_{32}|\approx0, &&\,\,\,\phi^0_{32}\approx -172.87^{\rm o},\,7.13^{\rm o}.
\label{ang32}
\end{eqnarray}
Recall that, by definition, the relation $\phi_{32}=(\varphi_3-\varphi_2)/2$ must hold for any acceptable pair of Majorana phases $\varphi_2$ and $\varphi_3$. This means that the hypothetical extraction of the values of the Majorana phases from the MMs $\mu_{21}^{\rm M}$ and $\mu_{31}^{\rm M}$ would automatically establish the value of the Majorana phase difference $\phi_{32}$, which is a parameter of $\mu_{32}^{\rm M}$. Nevertheless, from the estimations given in Ecs.~(\ref{ang21}), (\ref{ang31}) and (\ref{ang32}), it is clear that $(\varphi^0_3-\varphi^0_2)/2\approx2.44^{\rm o}\ne\phi^0_{32}$. %Imagine a scenario in which $\varphi_{2}=\varphi_2^0$ and $\varphi_{3}=\varphi_3^0$, so that $\mu^{\rm M}_{21}$ and $\mu^{\rm M}_{31}$ vanish. What our estimation of the saddle points $\varphi_2$, $\varphi_3$ and $\phi_{32}^0$ tells us is that the relation between the Majorana phases and the phase difference $\phi_{32}$ is not compatible with a situation in which $\phi_{32}=\phi_{32}^0$, which would eliminate $\mu^{\rm M}_{32}$ as well.
In other words, if $CP$ is violated by PMNS phases, at least one neutrino MM must have a nonzero value. An analogous situation happens when one observes Ecs.~(\ref{muest1}), (\ref{muest2}) and (\ref{muest3}), according to which $\phi^{\rm max}_{32}$ si very different to $(\varphi^{\rm max}_3-\varphi^{\rm max}_2)/2$. A similar reasoning indicates that no more than two MMs can  be maximal.

\subsection{CKM--like mixing}
\begin{figure}[!ht]
\center
\includegraphics[width=13cm]{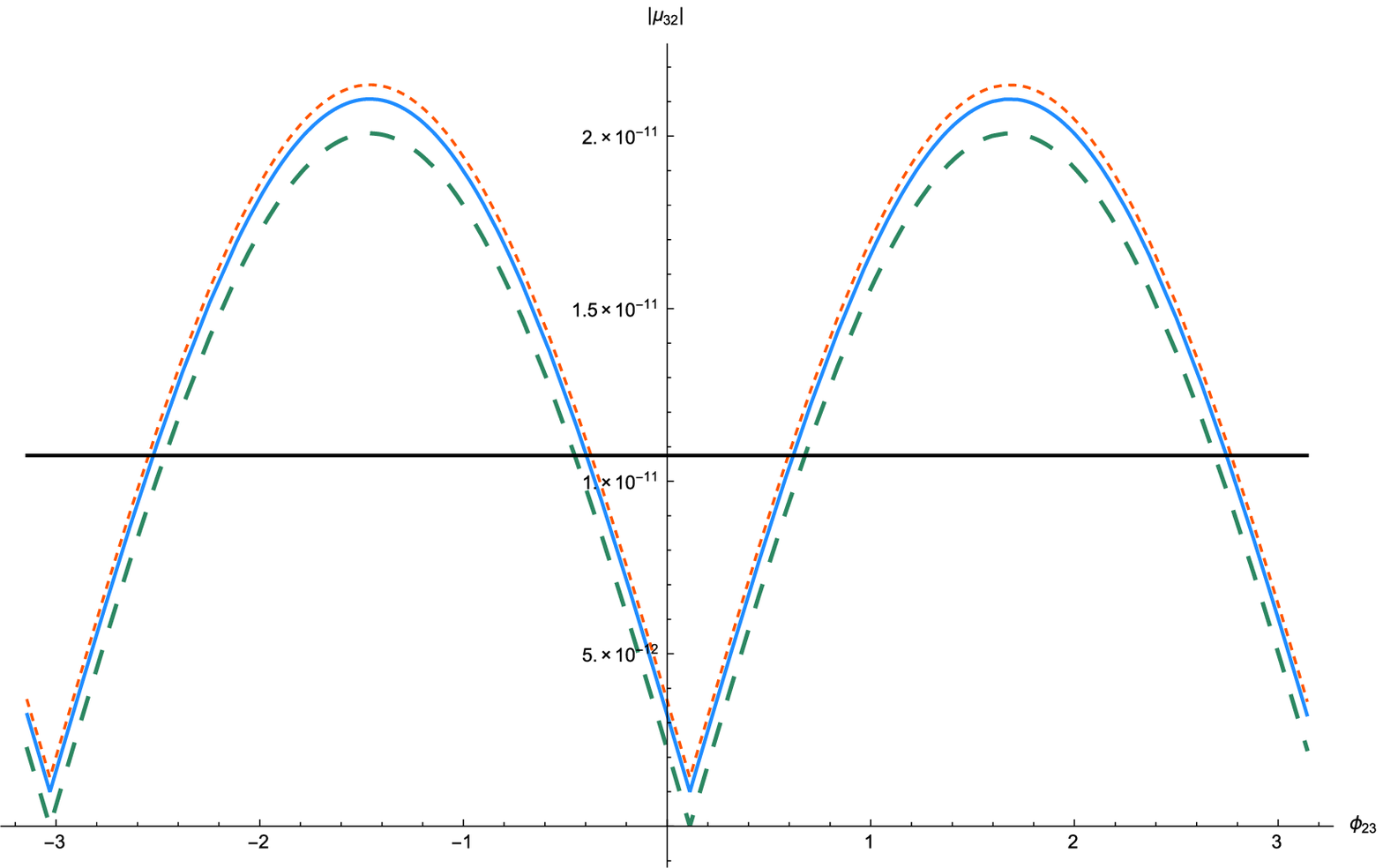}
\includegraphics[width=3cm]{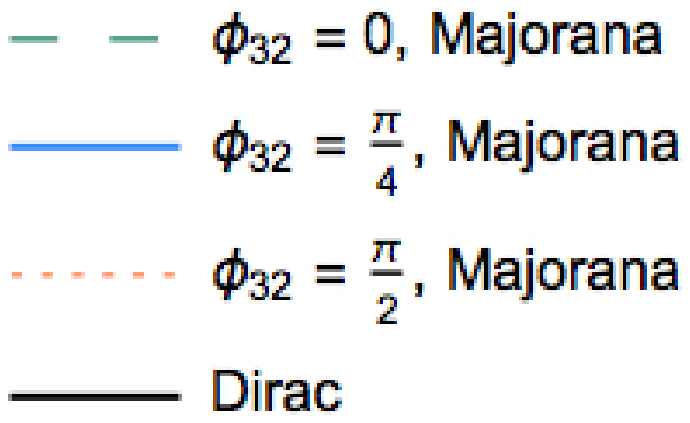}
\caption{\label{phi32} Allowed regions for $|\mu_{32}|$, as a function of $\phi_{23}$, for $\delta=-\pi/2$ and different Majorana phase differences $\phi_{32}$, in a scenario characterized by a CKM--like right--lepton mixing. The upper bound on the contributions to the corresponding MM of Dirac type, also shown in this figure, is represented by the horizontal line.}
\end{figure}

In general, the right PMNS matrix, ${\cal R}$, can be parametrized as~\cite{GK}
\begin{eqnarray}
{\cal R}=
\left(
\begin{array}{ccc}
c_{12}\,c_{13} & s_{12}\,c_{13} & s_{13}\,e^{-i\delta_{\rm R}}
\\
-s_{12}\,c_{23}\,-c_{12}\,s_{23}\,s_{13}\,e^{i\delta_{\rm R}} & c_{12}\,c_{23}-s_{12}\,s_{23}\,s_{13}\,e^{i\delta_{\rm R}} & s_{23}\,c_{13}
\\
s_{12}\,s_{23}-c_{12}\,c_{23}\,s_{13}e^{i\delta_{\rm R}} & -c_{12}\,s_{23}-s_{12}\,c_{23}\,s_{13}\,e^{i\delta_{\rm R}} & c_{23}\,c_{13}
\end{array}
\right),
\end{eqnarray}
where the sine and the cosine of the right mixing angles are respectively denoted by $s_{jk}=\sin\theta^{\rm R}_{jk}$, $c_{jk}=\cos\theta^{\rm R}_{jk}$, and $\delta_{\rm R}$ is the right Dirac phase. Assuming that the mixing angles $\theta^{\rm R}_{jk}$ are very close to 0 or to $\pi$, the off--diagonal terms are suppressed and the role played by the right Dirac phase $\delta_{\rm R}$ becomes marginal. In such context, ${\cal R}$ is very close to the identity matrix and is almost real.
%In this context, we write the $\mu^{\rm M}_{\beta\alpha}$ as
%\begin{eqnarray}
%\mu^{\rm M}_{\beta\alpha}&\approx& i\mu_{\rm B}\frac{g_Lg_R}{8\pi^2}\sin\zeta\,\cos\zeta\,m_e\frac{m_2^2-m_1^2}{m_2^2m_1^2}\Big| q_{1,\alpha\beta}\,\sin(\omega+\phi_{\beta\alpha})
%-q_{1,\beta\alpha}\,\sin(\omega+\phi_{\alpha\beta})+q_{2,\alpha\beta}\,\sin(\omega+\phi_{\beta\alpha}+\delta)
%\nonumber \\ \nonumber \\ &&
%-q_{2,\beta\alpha}\,\sin(\omega
%+\phi_{\alpha\beta}+\delta)+q_{3,\alpha\beta}\,\sin(\omega+\phi_{\beta\alpha}-\delta)-q_{3,\beta\alpha}\,\sin(\omega+\phi_{\alpha\beta}-\delta) \Big|.
%\end{eqnarray}
%The definitions of the coefficients $q_{j,\beta\alpha}$, in terms of , are shown in the Appendix.
Assuming that $\omega=0$, we find
\begin{eqnarray}
|\mu^{\rm M}_{\beta\alpha}|&\approx&\mu_{\rm B}\frac{g_Lg_R}{8\pi^2}\sin\zeta\,\cos\zeta\,m_e\frac{m_2^2-m_1^2}{m_2^2m_1^2}\left| c_{e,\beta\alpha}{\cal R}_{e1}+c_{\mu,\beta\alpha}{\cal R}_{\mu 2}+c_{\tau,\beta\alpha}{\cal R}_{\tau 3} \right|
\nonumber \\ \nonumber \\ &&
\lesssim\mu_{\rm B}\,(2\times10^{-11}{\rm GeV}^{-1})\left(\, |c_{e,\beta\alpha}|+|c_{\mu,\beta\alpha}|+|c_{\tau,\beta\alpha}| \,\right).
\end{eqnarray}
The whole set of coefficients $c_{j,\beta\alpha}$ is provided in Appendix~\ref{ccoefsdef}. Each $\mu^{\rm M}_{\beta\alpha}$ involves one $c_{j,\beta\alpha}$ coefficient that is equal to zero and two coefficients which are not. Each nonzero coefficient depends on only one charged--lepton mass, to which it is proportional, so that any MM involves only two different charged--lepton masses. Dependence on the left Dirac phase $\delta$ and on two Majorana phase differences, $\phi_{\beta\alpha}$ and $\phi_{\alpha\beta}$, is also present.
However, any MM is mostly sensitive to only one of such Majorana phase differences. This feature is dictated by the largest charged--lepton mass upon which a given MM depends.
To illustrate this point, consider the case of the $\mu^{\rm M}_{21}$ MM, which involves the coefficients $c_{e,21}\propto m_e$, $c_{\mu,21}\propto m_\mu$, and $c_{\tau,21}=0$.
%the $c_{1,\beta\alpha}$ are given by
%\begin{eqnarray}
%c_{1,21}&=&m_e\,\cos\theta_{13}\,\sin\theta_{12}\,\sin\phi_{21},
%\\ \nonumber \\
%c_{2,21}&=&m_\mu\left[ \cos\theta_{23}\,\sin\theta_{12}\,\sin\phi_{12}-\cos\theta_{12}\,\sin\theta_{13}\,\sin\theta_{23}\,\sin(\delta-\phi_{12}) \right]
%\\ \nonumber \\
%c_{2,31}&=&0.
%\end{eqnarray}
Since the muon mass is much larger than the electron mass, the contribution enclosed by $c_{e,21}$ is suppressed with respect to $c_{\mu,21}$. Noting that the Majorana phase difference $\phi_{21}$ only appears in the $c_{e,21}$ coefficient, it is clear that the magnitude of this MM is mainly sensitive to the phase difference $\phi_{12}$. The Dirac phase, being involved in the dominant $c_{\mu,21}$ coefficient also plays a role. This dominance of only one Majorana phase difference is even more dramatic in the case of $\mu_{32}^{\rm M}$, which depends on the masses of the electron and the tau lepton. Since here the ratio between the charged--lepton masses is an order of magnitude larger than the in previous case, the effect of the $\phi_{31}$ phase is even more suppressed. The case of the remaining MM, $\mu^{\rm M}_{32}$, is a little bit more equilibrated, for the difference between $m_\mu$ and $m_\tau$ is the smallest among all the differences of masses of charged leptons.
The behavior of this MM, as a function of $\phi_{23}$, is illustrated in Fig.~(\ref{phi32}),
where the magnitude $|\mu^{\rm M}_{32}|$ has been plotted for $\delta=-\pi/2$ and different choices of the Majorana phase difference $\phi_{32}$. This MM is mostly sensitive to the $\phi_{23}$ phase difference, but the effects of the phase difference $\phi_{32}$ can be appreciated in this figure. This contrasts with the situation of the other two MMs, whose graphs would look practically the same for any value of the subdominant Majorana phase differences. Additionally, Fig.~(\ref{phi32}) shows the upper bound on the contributions to the Dirac--like MM $\mu_{32}^{\rm D}$, which is represented by a horizontal line at $|\mu_{32}^{\rm D}|\approx1.07\times10^{-11}\mu_{\rm B}$.

Using the value $\delta=-\pi/2$, for the Dirac phase, yields the following upper bounds on the neutrino MMs:
\begin{eqnarray}
|\mu_{21}^{\rm M}|\lesssim&8.89\times10^{-13}\,\mu_{\rm B}, &
\left\{
\begin{array}{ll}
\phi^{\rm max}_{12}\approx102.00^{\rm o},-78.00^{\rm o}, \,\,\, r_2^{\rm max}\approx156^{\rm o},
\\ \\
\phi^{\rm max}_{21}=\pm90^{\rm o},  \,\,\,l_2^{\rm max}=180^{\rm o},
\end{array}
\right.
\label{-13MM}
\\ \nonumber \\
|\mu^{\rm M}_{31}|\lesssim&1.36\times10^{-11}\,\mu_{\rm B}, &
\left\{
\begin{array} {ll}
\phi^{\rm max}_{13}\approx75.59^{\rm o}, -104.41^{\rm o}, \,\,\,r_3^{\rm max}\approx-151.18,
\\ \\
\phi^{\rm max}_{31}=0,\,\,180^{\rm o}, \,\,\, l_3^{\rm max}=0^{\rm o},
\end{array}
\right.
\\ \nonumber \\
|\mu^{\rm M}_{32}|\lesssim&2.15\times10^{-11}\,\mu_{\rm B}, &
\left\{
\begin{array}{ll}
\phi^{\rm max}_{23}\approx96.40^{\rm o},-83.60^{\rm o}, &
\\ \\
\phi^{\rm max}_{32}=\pm90^{\rm o}.
\end{array}
\right.
\end{eqnarray}
In the last three equations, we have provided the phase differences $\phi_{\beta\alpha}^{\rm max}$ yielding these maxima, and the Majorana phases associated to such differences. From Eq.~(\ref{-13MM}), a difference among the maximal right mixing, discussed in the last subsection, and the CKM--like right mixing can be pointed out: the upper bound on the magnitude of the $\mu^{\rm M}_{21}$ is more than one order of magnitude larger in the former case than in the latter. This means that if experiments measured a $\mu_{21}^{\rm M}$ within the range $10^{-11}-10^{-12}$, the scenario with maximal right mixing ${\cal L}={\cal R}$ would be favored, with respect to the case of CKM--like mixing, in the context of the left--right model considered in the present work.
\\

Similarly to what occurred in the context of the maximal neutrino mixing that we discussed in the previous subsection, here each neutrino transition MM $\mu_{\beta\alpha}^{\rm M}$ vanish for certain Majorana phase differences $\phi^0_{\beta\alpha}$ and $\phi^0_{\alpha\beta}$. For $\delta=-\pi/2$, we find the following sets of angles:
\begin{eqnarray}
|\mu^{\rm M}_{21}|\approx0, &&
\left\{
\begin{array}{l}
\phi^0_{12}\approx12.00^{\rm o},-168.00^{\rm o}, \,\,\,r_2^0\approx-24^{\rm o},
\\ \nonumber \\
\phi_{21}^0=0^{\rm o},\,180^{\rm o},\,\,\,l_2^0=0^{\rm o},
\end{array}
\right.
\\ \nonumber \\
|\mu^{\rm M}_{31}|\approx0, &&
\left\{
\begin{array}{l}
\phi^0_{13}\approx165.59^{\rm o},-14.41^{\rm o},\,\,\, r_3^0\approx28.82^{\rm o},
\\ \nonumber \\
\phi^0_{31}=\pm90^{\rm o},\,\,\, l_3^0=180^{\rm o},
\end{array}
\right.
\\ \nonumber \\
|\mu^{\rm M}_{32}|\approx0,&&
\left\{
\begin{array}{l}
\phi^0_{23}\approx6.40^{\rm o},-173.60^{\rm o},
\\ \nonumber \\
\phi^0_{32}=0^{\rm o},\,180^{\rm o}.
\end{array}
\right.
\end{eqnarray}
Keeping in mind these results, visualize a setting in which we have Majorana phase differences $\phi_{21}=\phi^0_{21}$, $\phi_{12}=\phi^0_{12}$, $\phi_{31}=\phi^0_{31}$, and $\phi_{13}=\phi^0_{13}$, so that $\mu^{\rm M}_{21}\approx0$ and $\mu^{\rm M}_{31}\approx0$. Then note that
\begin{eqnarray}
\frac{l^0_2-r^0_3}{2}&\approx&-14.41^{\rm o}\ne\phi_{23}^0,
\\ \nonumber \\
\frac{l_3^0-l_2^0}{2}&\approx&102^{\rm o}\ne\phi^0_{32}.
\end{eqnarray}
We find, then, that there cannot be a set of Majorana phase differences that make all three MMs equal to zero and satisfy the equations $\phi^0_{23}=(l_2^0-r_3^0)/2$ and $\phi^0_{32}=(l_3^0-r_2^0)/2$, at the same time, so that two vanishing neutrino MMs require the third one to be nonzero. This means that, in the presence of $CP$ violation driven by Majorana and Dirac phases, there must be, at least, one nonzero neutrino MM. Finally, as it happened in the case ${\cal L}={\cal R}$, we observe that optimal sets of Majorana phase differences cannot generate more than two maximal MMs.
\\

\section{Conclusions}
\label{conc}
In this paper, we have studied the  magnetic moments of neutrinos living in a world in which non--manifest left--right symmetry, lying beyond the Standard Model, describes nature at a high--energy scale. The resulting contributions arose from the calculation, performed at the one--loop level, of the neutrino electromagnetic vertex $\gamma\nu\nu$, for which we took into account the role of charged currents featuring light and heavy mixed charged gauge bosons that couple to Majorana neutrinos. We derived an expression for the magnetic moments, of which only off--diagonal terms are nonzero. This is a characteristic of Majorana neutrinos, as the ones living within this description of new physics. We carried out the calculation in the general $R_\xi$--gauge and, in the limit corresponding to the unitary gauge, we found agreement with previous investigations. However, we went further and expressed our result explicitly in terms of the $CP$--violating phases that are part of the parametrization of neutrino mixing. The full set of neutrino magnetic moments can be seen as entries of a $3\times3$ hermitian matrix, which turns out to be antisymmetric. This reduces the number of independent magnetic moments to just three. Since we dealt with Majorana neutrinos, there were three $CP$--violating phases: one Dirac and two Majorana. We considered two manners in which right--handed neutrino mixing may be realized. In the first place, we examined a maximal right neutrino mixing in which the right mixing matrix coincides with the one describing the mixing of left--handed neutrinos. Then, we considered a right--handed neutrino mixing matrix in which off--diagonal terms are suppressed with respect to those in the diagonal. Though we attenuated the effects from off--diagonal terms by assuming right mixing angles that are close to 0 or $\pi$, we left the $CP$ Majorana and Dirac phases free of assumptions. Then we discussed the effect of $CP$ violation, driven by these phases, and found that even in the case that Dirac $CP$--violation is present, certain values of the Majorana phases may attenuate the contributions, and even eliminate them. Nevertheless, we have pointed out that at least one of the neutrino magnetic moments must be nonzero. Similarly, we found that a no more than two neutrino magnetic moments can have a maximum value. Estimations show that the MMs are, at most, of order $10^{-11}\mu_{\rm B}$.

\begin{acknowledgments}
We particularly want to thank Dr. Juan Barranco for usefull discussions.
We acknowledge financial support from CONACYT and SNI (M\'exico). D. D. is  grateful to Conacyt (M\'exico) S.N.I. and Conacyt project (CB-156618), DAIP
project (Guanajuato University) and PIFI (Secretaria de Educacion
Publica, M\'exico) for financial support.
\end{acknowledgments}

\appendix

\section{Definitions of coefficients $\boldsymbol{m_{j,\beta\alpha}}$}
\label{mcoefs}
The coefficients $m_{j,\beta\alpha}$, used in Eqs.~(\ref{mdm2}), are defined as
\begin{eqnarray}
m_{1,11}&=&\cos ^2\theta _{12} \left(m_e
   \cos ^2\theta _{13}+\sin
   ^2\theta _{13} \left(m_{\mu }\sin
   ^2\theta _{23}+ m_{\tau
   }\cos
   ^2\theta _{23}\right)\right)+\sin ^2\theta
   _{12} \left( m_{\mu }\cos ^2\theta
   _{23}+ m_{\tau }\sin ^2\theta
   _{23}\right),
\\ \nonumber \\
m_{1,12}&=&\sin \theta _{12} \cos
   \theta _{12}\, \left(m_e \cos
   ^2\theta _{13}+\sin
   ^2\theta _{23} \left( m_{\mu
   }\sin
   ^2\theta _{13}-m_{\tau }\right)+\cos ^2\theta
   _{23} \left(m_\tau\sin ^2\theta
   _{13}-m_{\mu
   }\right)\right),
\\ \nonumber \\
m_{1,13}&=&\sin \theta _{12} \sin
   \theta _{23} \cos\theta _{13} \cos\theta _{23}\left(m_{\tau
   }-m_{\mu }\right),
\\ \nonumber \\
m_{1,22}&=&\sin ^2\theta _{12} \left(m_e
   \cos ^2\theta _{13}+\sin
   ^2\theta _{13} \left( m_{\mu }\sin
   ^2\theta _{23}+ m_{\tau
   }\cos
   ^2\theta _{23}\right)\right)+\cos ^2\theta
   _{12} \left( m_{\mu }\cos ^2\theta
   _{23}+ m_{\tau }\sin ^2\theta
   _{23}\right),
\\ \nonumber \\
m_{1,23}&=&\sin \theta _{23}\cos\theta _{12} \cos\theta _{13} \cos\theta _{23}\left(m_{\mu
   }-m_{\tau }\right),
\\ \nonumber \\
m_{1,33}&=&m_e \sin ^2\theta _{13}+\cos
   ^2\theta _{13} \left( m_{\mu }\sin
   ^2\theta _{23}+ m_{\tau
   }\cos
   ^2\theta _{23}\right),
\\ \nonumber \\
m_{2,11}&=&\sin \theta _{12} \sin\theta _{13} \sin\theta _{23} \cos\theta _{12} \cos\theta _{23} \left(m_{\mu
   }-m_{\tau }\right)
\\ \nonumber \\
m_{2,12}&=&\sin \theta _{13} \sin\theta _{23} \cos
   ^2\theta _{12} \cos\theta _{23} \left(m_{\tau
   }-m_{\mu }\right),
\\ \nonumber \\
m_{2,13}&=&\sin \theta _{13} \cos\theta _{12} \cos\theta _{13} \left(m_e- m_{\mu }\,\sin
   ^2\theta _{23}-m_{\tau
   }\,\cos
   ^2\theta _{23} \right),
\\ \nonumber \\
m_{2,21}&=&\sin ^2\theta _{12} \sin\theta _{13} \sin\theta _{23} \cos\theta _{23} \left(m_{\mu
   }-m_{\tau }\right),
\\ \nonumber \\
m_{2,22}&=&\sin\theta _{12} \sin\theta _{13} \sin\theta _{23} \cos\theta _{12} \cos\theta _{23} \left(m_{\tau
   }-m_{\mu }\right),
\\ \nonumber \\
m_{2,23}&=&\sin \theta _{12} \sin\theta _{13} \cos\theta _{13} \left(m_e-m_{\mu }\,\sin
   ^2\theta _{23} -m_{\tau
   }\,\cos
   ^2\theta _{23}\right),
\end{eqnarray}
Additionally, $m_{1,\beta\alpha}=m_{1,\alpha\beta}$ and $m_{2,3\alpha}=0$.

\section{Definitions of coefficients $\boldsymbol{c_{j,\beta\alpha}}$}
\label{ccoefsdef}
The coefficients $c_{j,\beta\alpha}$ are antisymmetric, that is, $c_{j,\beta\alpha}=-c_{j,\alpha\beta}$. Their explicit expressions are
\begin{eqnarray}
c_{e,21}&=&m_e\,\cos\theta_{13}\,\sin\theta_{12}\,\sin\phi_{21},
\\ \nonumber \\
c_{\mu,21}&=&m_\mu(\cos\theta_{23}\,\sin\theta_{12}\,\sin\phi_{12}+\cos\theta_{12}\,\sin\theta_{13}\sin\theta_{23}\,\sin(\delta+\phi_{12}),
\\ \nonumber \\
c_{\tau,21}&=&0,
\\ \nonumber \\
c_{e,31}&=&-m_e\,\sin\theta_{13}\,\sin(\delta-\phi_{31}),
\\ \nonumber \\
c_{\mu,31}&=&0,
\\ \nonumber \\
c_{\tau,31}&=&m_\tau(\cos\theta_{12}\,\cos\theta_{23}\sin\theta_{13}\,\sin(\delta+\phi_{13})-\sin\theta_{12}\,\sin\theta_{23}\,\sin\phi_{13}),
\end{eqnarray}
\begin{eqnarray}
c_{e,32}&=&0,
\\ \nonumber \\
c_{\mu,32}&=&m_\mu\,\cos\theta_{13}\,\sin\theta_{23}\,\sin\phi_{32},
\\ \nonumber \\
c_{\tau,32}&=&m_\tau(\cos\theta_{12}\,\sin\theta_{23}\,\sin\phi_{23}+\cos\theta_{23}\,\sin\theta_{12}\,\sin\theta_{13}\,\sin(\delta+\phi_{23})).
\end{eqnarray}

\end{document}